\documentclass[sigconf,10pt]{acmart}

\setcopyright{rightsretained}

\usepackage{subcaption}
\usepackage[capitalise,noabbrev]{cleveref}

\usepackage{amsmath}
\usepackage{nccmath}

\newcommand{\heading}[1]{\vspace{4pt}\noindent\textbf{#1.}}

\acmDOI{10.475/123_4}

\acmISBN{978-1-4503-7020-2}

\acmConference[HotNets'21]{The 20th ACM Workshop on Hot Topics in
Networks}{November 10-12, 2021}{Virtual Event, UK}
\acmYear{2021}
\copyrightyear{2021}

\begin{document}

\title[Tradeoffs in network monitoring]{{Towards a Cost vs. Quality Sweet Spot for Monitoring Networks}}

\vspace{-2mm}
\author{
Nofel Yaseen$^{\ddag\diamond}$,  Behnaz Arzani$^\ddag$, Krishna Chintalapudi$^\ddag$,  Vaishnavi Ranganathan$^\ddag$,\\
Felipe Frujeri$^\ddag$, Kevin Hsieh$^\ddag$, Daniel Berger$^\ddag$, Vincent Liu$^\diamond$, Srikanth Kandula$^\ddag$
}
\vspace{-2mm}
\affiliation{Microsoft$^\ddag$ and University of Pennsylvania$^\diamond$}
\renewcommand{\shortauthors}{N. Yaseen et al.}










\begin{abstract}

Continuously monitoring a wide variety of performance and fault metrics has become a crucial part of operating large-scale datacenter networks. In this work, we ask whether we can reduce the costs to monitor -- in terms of collection, storage and analysis -- by judiciously controlling how much and which measurements we collect.  By positing that we can treat almost all measured signals as sampled time-series, we show that we can use signal processing techniques such as the Nyquist-Shannon theorem to avoid wasteful data collection. We show that large savings appear possible by analyzing tens of popular measurements from a production datacenter network. We also discuss the technical challenges that must be solved when applying these techniques in practice.

\end{abstract}

\maketitle

\section{Introduction}
\label{sec:intro}

\begin{figure}
\centering
\includegraphics[width=\linewidth]{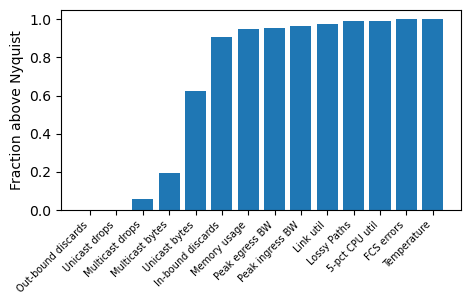}\vspace{-0.2in}
\caption{The fraction of devices~(collection points) at which our production data center currently measures various metrics above the Nyquist rate; each bar coalesces information from O($10^3$) devices.
\label{fig::intro}}\vspace{-0.1in}
\end{figure}

High availability guarantees are etched into the service-level agreements of data centers, and any failure in their adherence results in significant monetary impact.
As a result, data centers deploy large-scale monitoring systems that continuously monitor various performance metrics to help quickly identify/predict service disruptions and alleviate them.
For the most common class of monitoring systems, typical operation entails periodically sampling operational parameters across various components of a data center (e.g., CPU temperatures, packet drop counts, and path latencies); operators collect and analyze these measurements in real-time or store them for later analysis.

At the scale of modern data centers, monitoring systems can use a sizable portion of valuable storage, network bandwidth, and CPU resources. Thus, efficiency is crucial for production monitoring systems. Furthermore, there is a fundamental trade-off between monitoring cost and quality-- for a large class of underlying measurements be they interface counters, latency of pings and traceroutes, or results from sketches,  better quality generally requires greater cost but the actual trade-off remains nebulous.

Today, the typical reaction to this dilemma is ad-hoc and rather conservative.
In our survey of measurements collected from many production monitoring systems of a large cloud provider, we observe admins typically choose to collect as much information as possible subject to some (arbitrarily set) resource constraints. They often cannot answer whether their ad-hoc'ly chosen measurement rate is adequate for a given metric or whether measuring more (or less) frequently will lead to better (or no worse) insights? In fact, admins often express concern that collecting less information could lead to missing out on important insights.

Information theory provides techniques that help quantify this trade-off.
For instance, the Nyquist rate of a signal is the minimum sampling rate that fully captures the information in the signal. Some technical conditions apply such as the signal being bounded in the frequency domain which holds by definition for many practical metrics; for instance, when measuring the temperature of CPUs, the underlying thermodynamics limit the maximum rate at which temperatures change. We discuss other complications in~\cref{sec:practicalconcerns}. 

A key contribution of this paper is the somewhat surprising result that we can treat a wide range of metrics in production data centers as digital signals and find appropriate measurement frequencies~(using nyquist rate) which are several orders of magnitude smaller than the actual rates at which these metrics are measured today in  production. This indicates we can achieve similar measurement quality at much smaller monitoring costs. \cref{fig::intro} shows operators are over-sampling a diverse set of metrics at a vast majority of measurement points (methodology in \cref{sec:casestudy}). 

We offer a method that given a metric trace can compute the Nyquist rate; key issues here are how to cope with measurement noise and quantization effects. Using this method we quantify the possible gains on many real-world metrics.

We note that computing the appropriate measurement rate aposteriori does not suffice: to reduce monitoring costs, we must use an appropriate measurement rate while measuring the metric of interest; also, the choice of measurement rate must be robust to variations in the metric value such as sudden changes and phase shifts. To this end we propose a  dynamic sampling method which can adapt to changes in the underlying signal's Nyquist rate in near real-time and outline the research questions that arise in using such a system in practice. We further show preliminary results demonstrating the effectiveness of this approach by comparing its output with that of existing monitoring systems (which operate at ad-hoc sampling rates chosen by network operators) in our technical report~\cite{arzani2021DSP}.


While signal processing techniques such as compressive sensing and sparse FFT have been applied before, to our knowledge, we are unaware of any prior work that applies the Nyquist principle to find appropriate measurement rates for datacenter metrics. Our initial results are promising and point to large untapped gains. Finally, we believe that treating datacenter measurements as signals can lead to fruitful future work since there is a vast literature on signal processing beyond the Nyquist principle.

\begin{figure}
\centering
\includegraphics[width=0.9\linewidth]{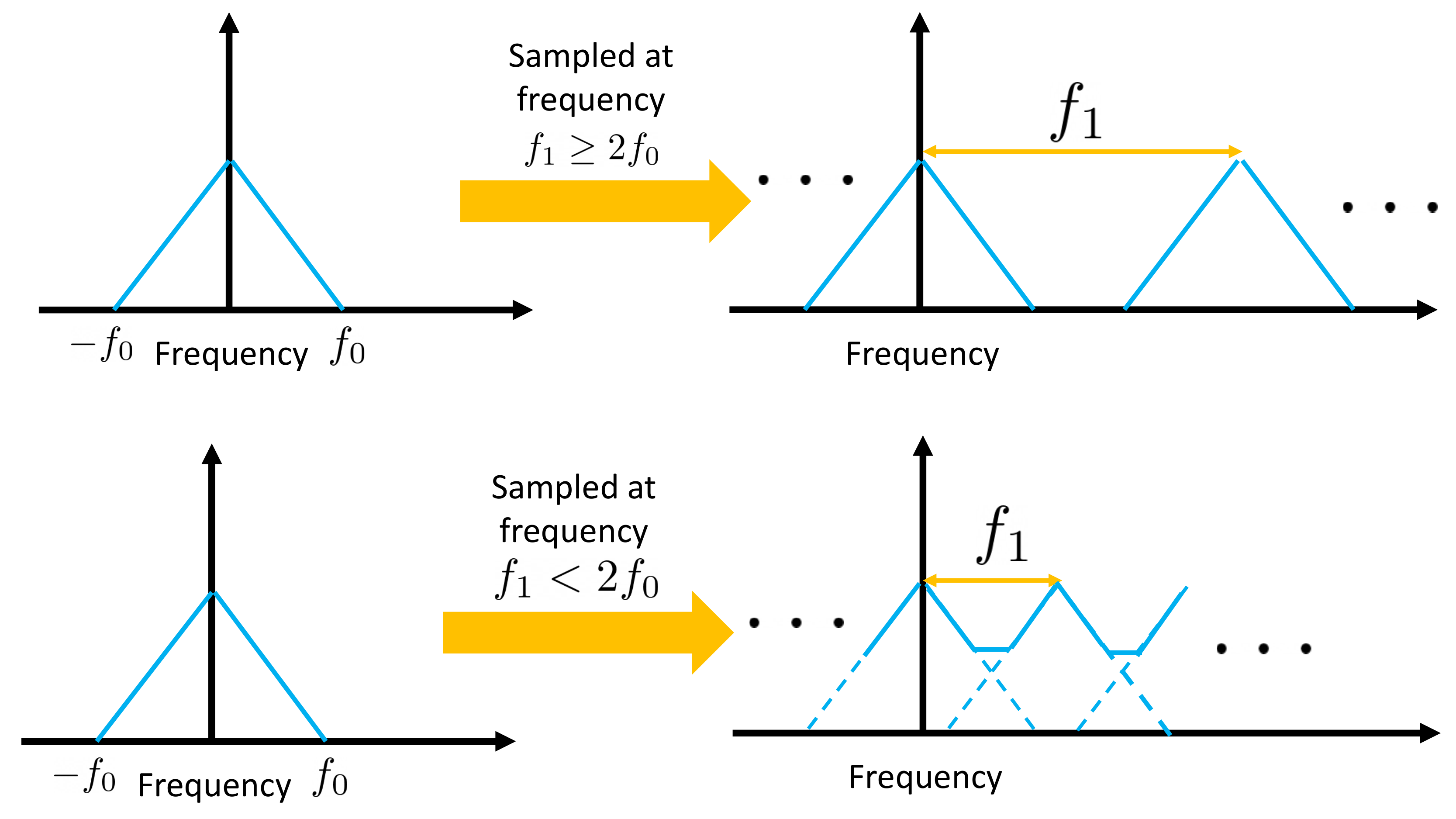}\vspace{-0.1in}
\caption{Showing the result, in the frequency domain, when sampling above and below the Nyquist rate. Sampling a signal at frequency $f_1$ and reconstructing it can be thought off, in the frequency domain, as adding copies of the signal which are $f_1$ apart.\label{fig::explanation}}\vspace{-0.05in} 
\end{figure}

\section{A Primer on Nyquist--Shannon}
\label{sec:primer}

Signals are functions of one or more independent variables, with the primary independent variable in most signals being time (though the theory applies equally to other variables). We can further divide these signals into two categories: continuous- and discrete-time signals where the difference is in the domain of the function.
More concretely, continuous-time signals are functions, $f(t)$, where $t: \mathbb{R} \rightarrow \mathbb{R}$, and discrete-time signals are functions, $f(T)$, where $T: \mathbb{N} \rightarrow \mathbb{R}$.
The outputs of today's measurement systems are discrete-time signals.
Sampling a signal converts continuous-time signals into discrete-time signals or down-samples discrete-time signals to reduce the costs to monitor and store telemetry.

\begin{figure*}[!t]
    \begin{subfigure}[b]{0.24\textwidth}
        \centering
        \includegraphics[width=\linewidth]{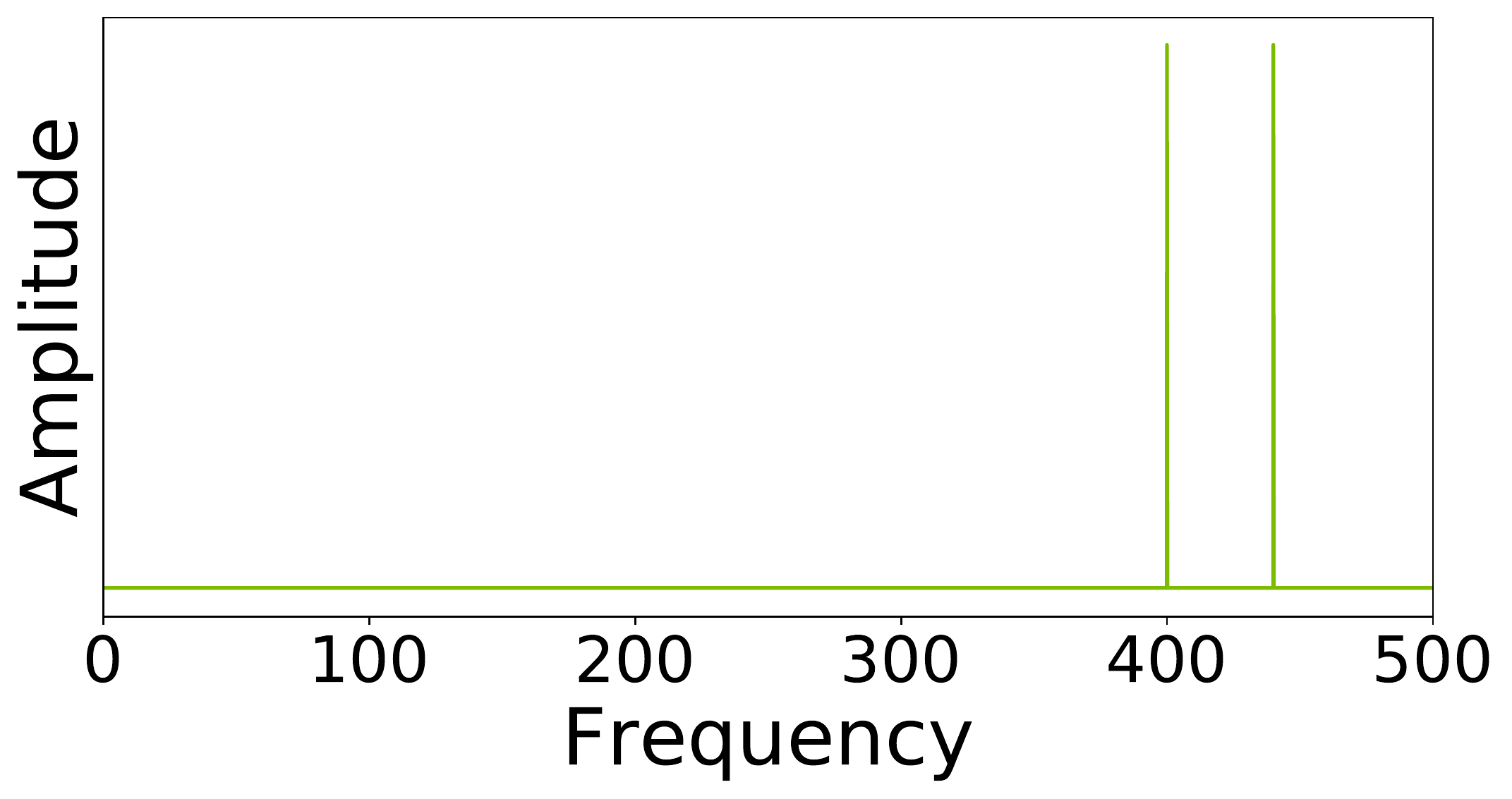}
        \caption{Original signal}
    \end{subfigure}
    \begin{subfigure}[b]{0.24\textwidth}
        \centering
        \includegraphics[width=\linewidth]{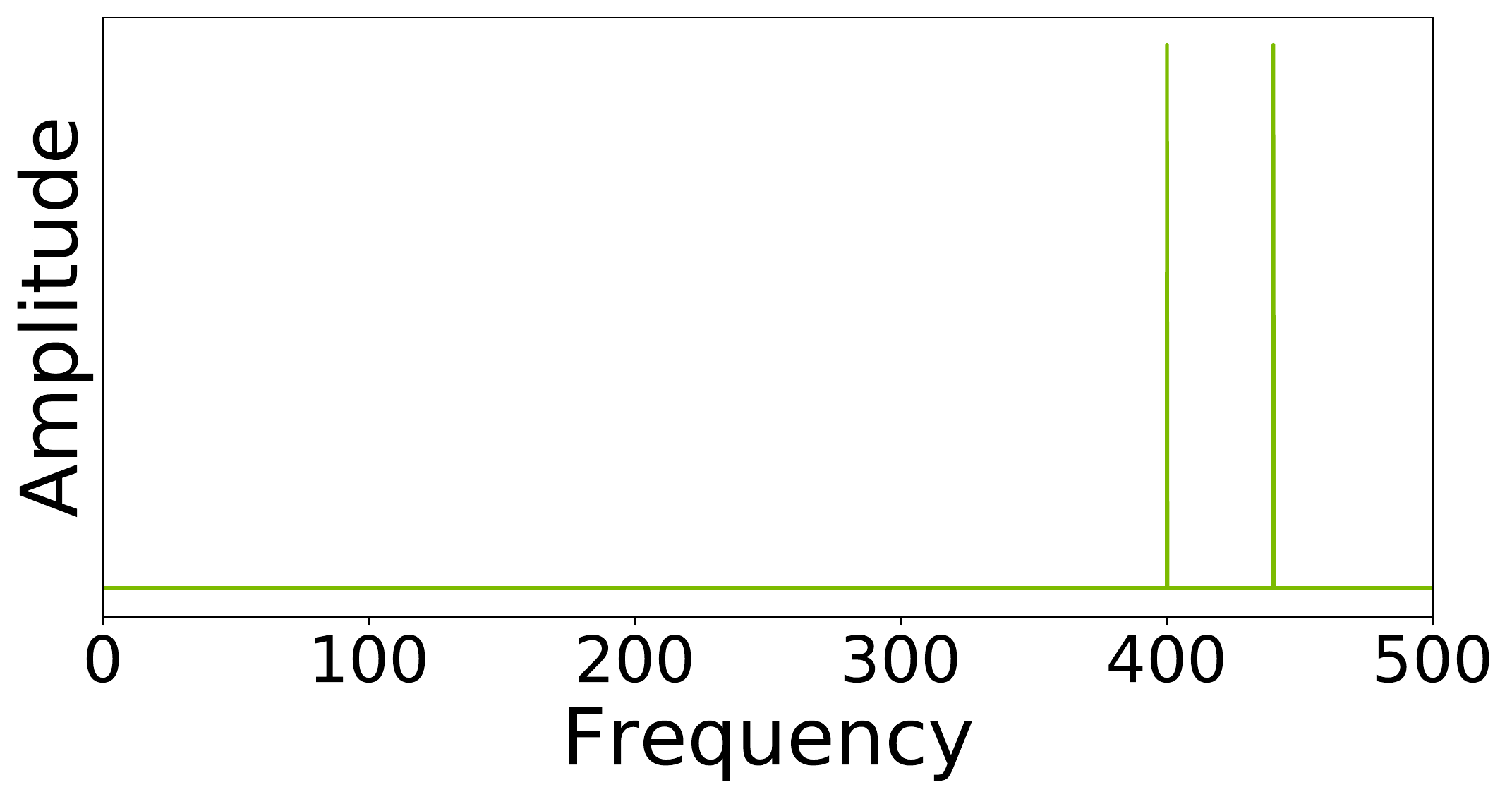}
        \caption{Sampled above Nyquist \label{f:3b}}
    \end{subfigure}
    \begin{subfigure}[b]{0.24\textwidth}
        \centering
        \includegraphics[width=\linewidth]{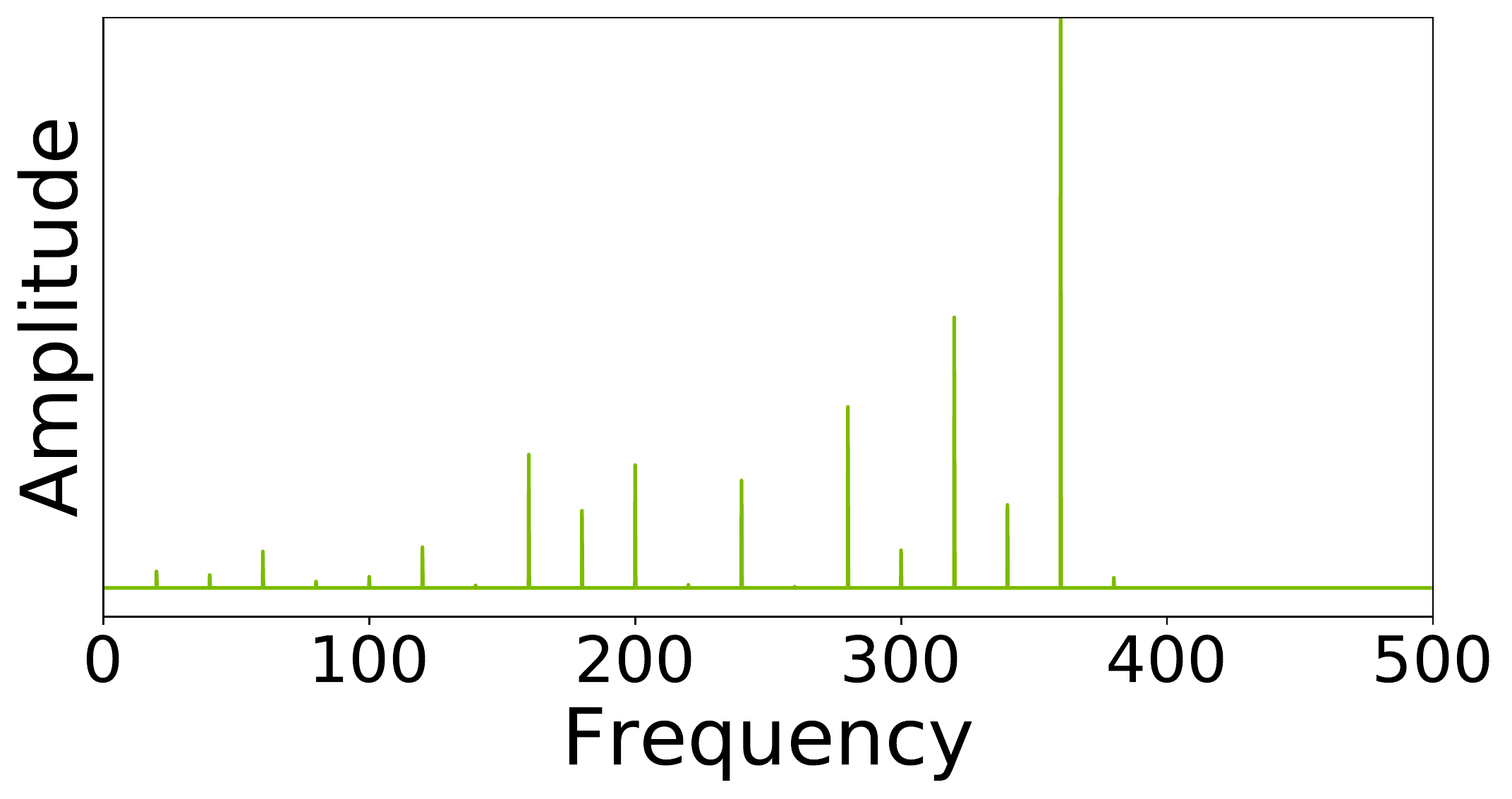}
        \caption{Sampled below Nyquist \label{f:3c}}
    \end{subfigure}
    \begin{subfigure}[b]{0.24\textwidth}
        \centering
        \includegraphics[width=\linewidth]{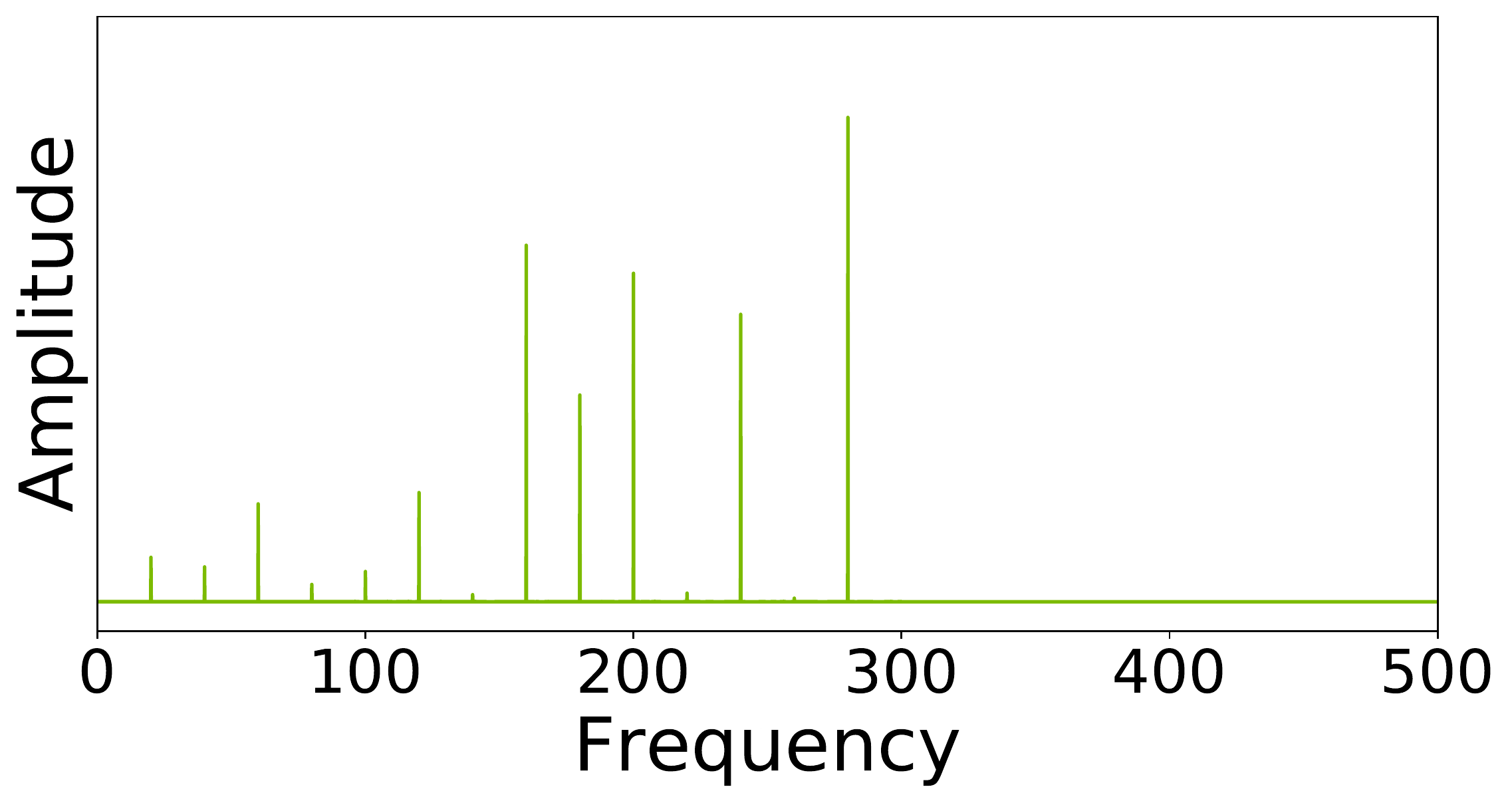}
        \caption{Sampled far below Nyquist\label{f:3d}}
    \end{subfigure}

    \begin{subfigure}[b]{0.24\textwidth}
        \centering
        \includegraphics[width=\linewidth]{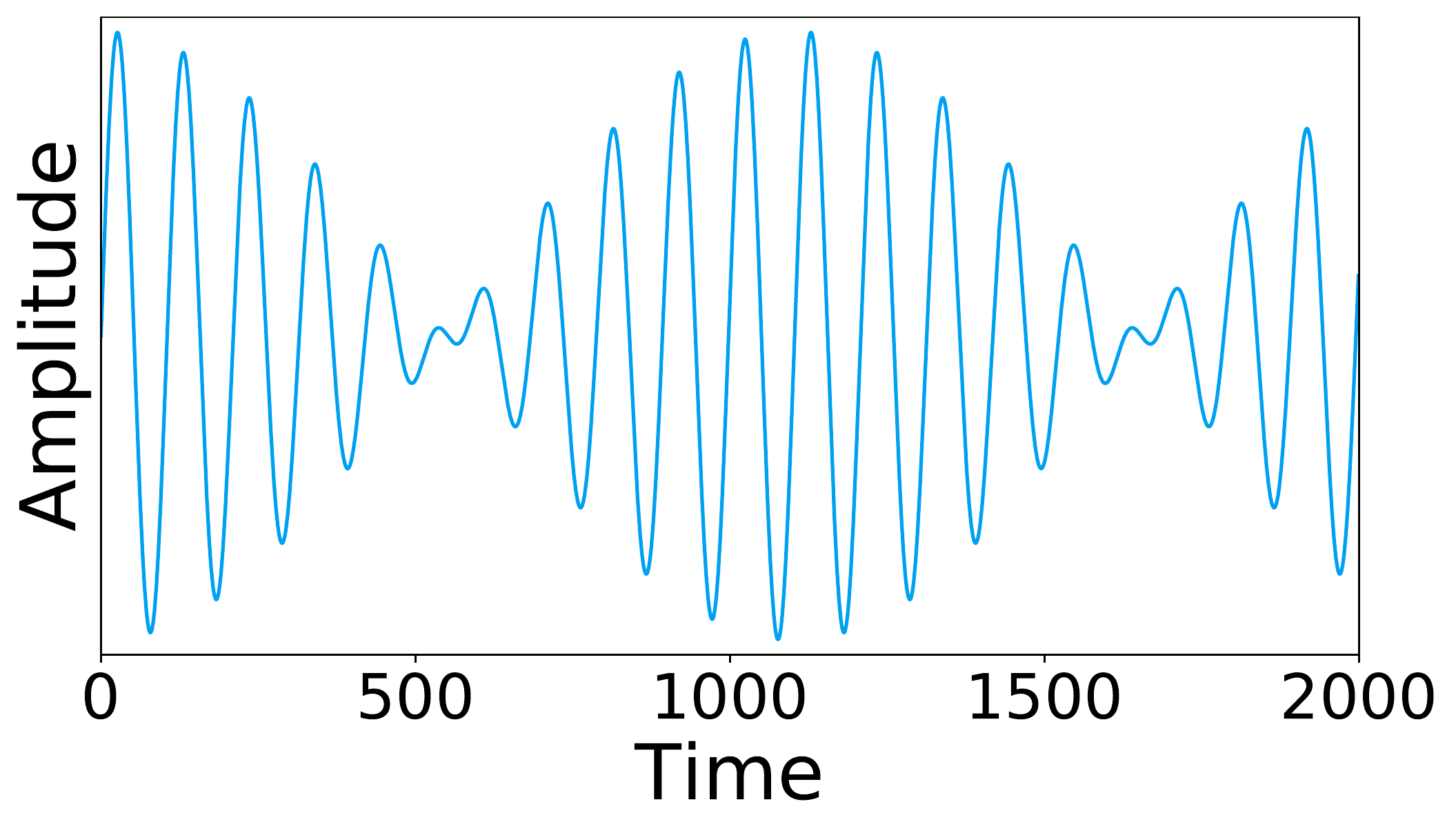}
        \caption{Original signal}
    \end{subfigure}
    \begin{subfigure}[b]{0.24\textwidth}
        \centering
        \includegraphics[width=\linewidth]{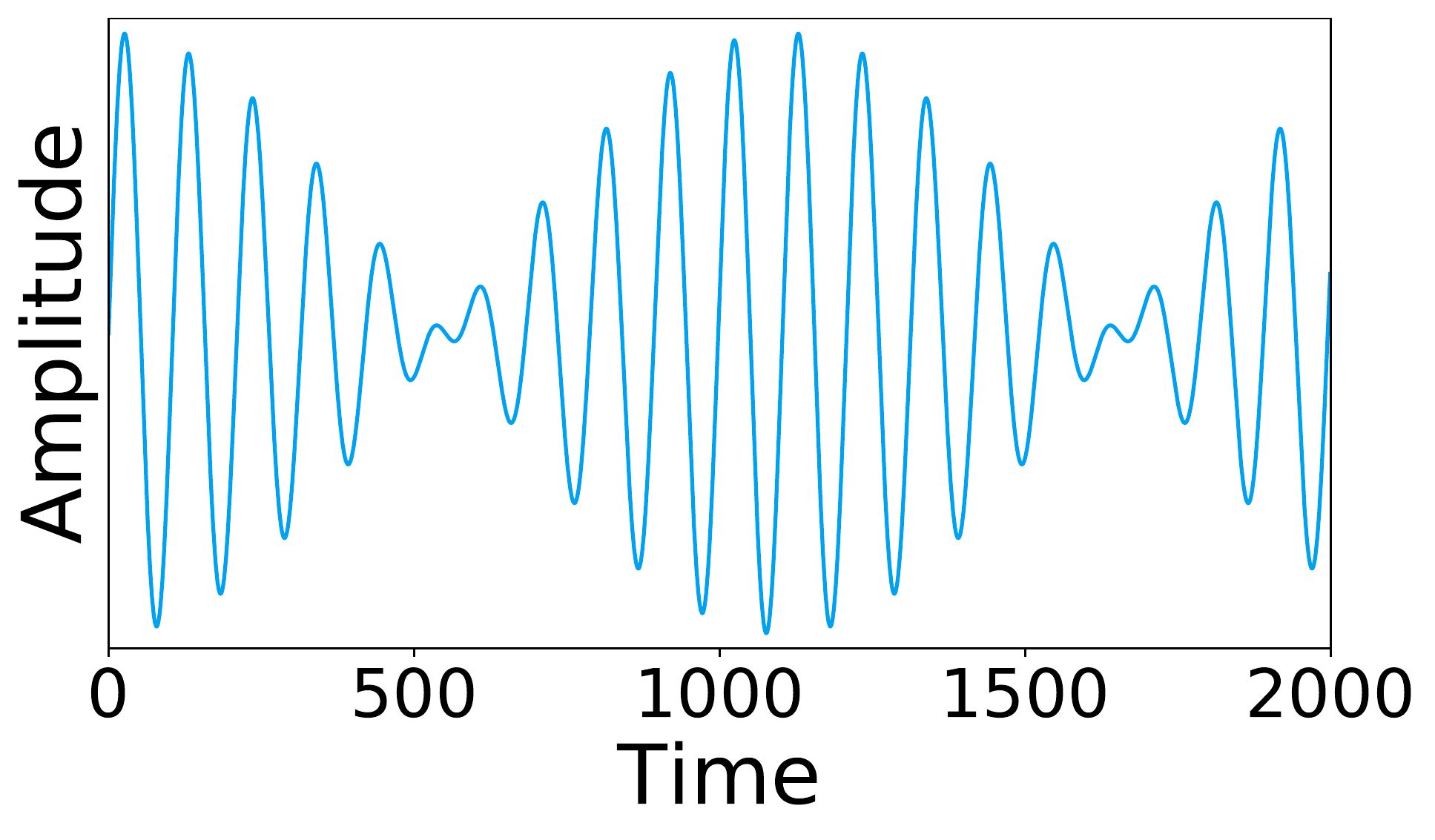}
        \caption{Reconstructed: ~\autoref{f:3b}}
    \end{subfigure}
    \begin{subfigure}[b]{0.24\textwidth}
        \centering
        \includegraphics[width=\linewidth]{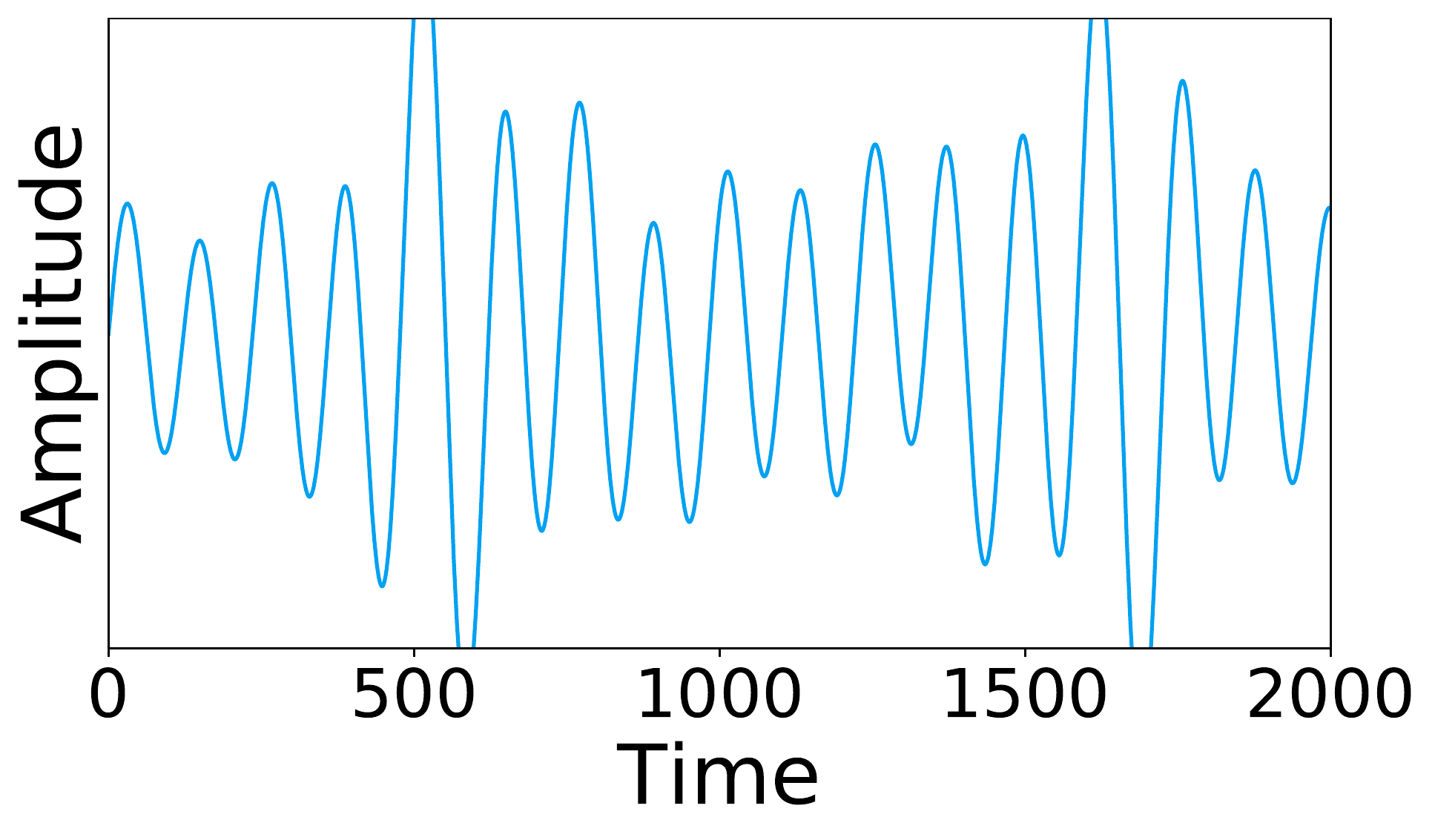}
        \caption{Reconstructed:~\autoref{f:3c}}
    \end{subfigure}
    \begin{subfigure}[b]{0.24\textwidth}
        \centering
        \includegraphics[width=\linewidth]{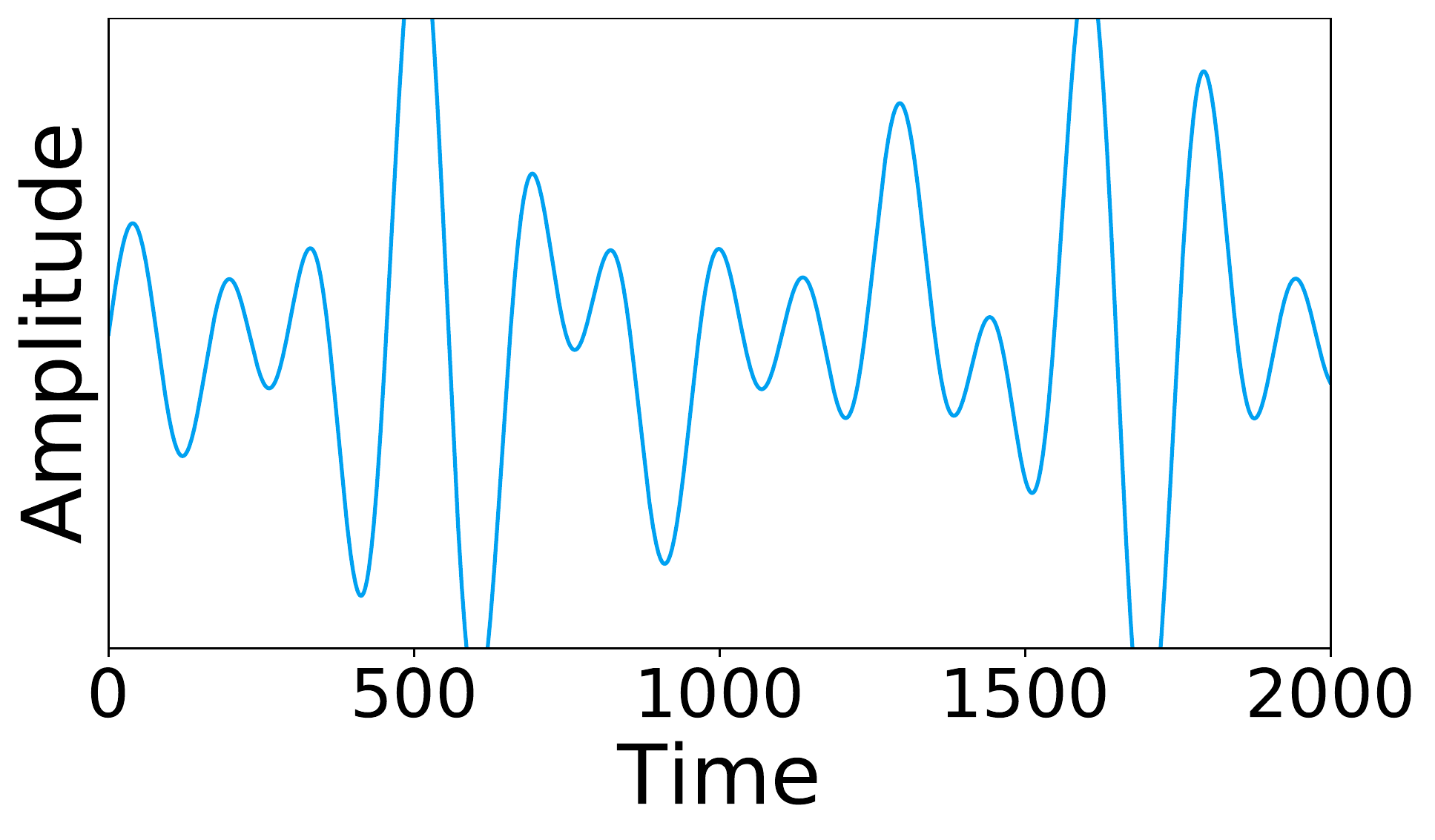}
        \caption{Reconstructed:~\autoref{f:3d}}
    \end{subfigure}
    \caption{
    A signal and its sampled versions in the frequency domain (top row).
    The original signal (bottom) is the superposition of two $\sin$ waves at $400$ and $440$\,Hz.
    The time-domain representations of the sampled versions are reconstructed and upsampled.
    Variants shown are: (a) the original signal, (b) the signal sampled at above the Nyquist rate ($890$ Hz), (c) the signal sampled at slightly below the Nyquist rate ($800$ Hz), and (d) the signal sampled at significantly below the Nyquist rate ($600$ Hz).
    Aliasing is observable in the frequency domain of (c) and (d).
    Reconstructing a signal (f) from the DFT of (d) results in a distorted result.
    \label{fig:background_nyquist}}
\end{figure*}

\heading{The Fourier transform}
One important property of signals is that they can be converted from functions over time into the frequency-domain (i.e., as mathematical functions of frequency). These conversions from the time to frequency domain are \textit{lossless}: we can recover the time-domain representation of the signal from its frequency domain representation and vice-versa.
The procedure to accomplish this translation is the Fourier transform, which produces a function whose magnitude at frequency $f$ is the amount of that frequency present in the original function.
An Inverse Fourier Transform converts the signal back to the time-domain.

The Fast Fourier Transform (FFT) is an efficient algorithm that applies over discrete sampled signals.
Given an input $N$ which can either be specified by the user or determined by the length of the signal, the FFT divides the frequency space from 0 to the maximum frequency in the signal into $N$ discrete bins and computes the signal power in each bin. Bin $j$ is computed as $j*(\text{sampling frequency}/N)$.
The square of these per-bin magnitudes is called the the Power Spectral Density (PSD); we will use this later in \cref{sec:casestudy} to analyze the Nyquist rate for deployed monitoring systems.

\heading{Nyquist rates and Fourier transforms in practice}
The \textit{Nyquist--Shannon theorem} states: if a function $x(t)$ contains no frequencies higher than $f_0$, then, sampling it at a rate at or above $2f_0$ ensures the original signal can be recovered completely; we call $2f_0$ the \textit{Nyquist rate} of $x(t)$~\cite{oppenheim1975digital}.

\autoref{fig::explanation} shows the effects of sampling at a rate above or below the Nyquist rate. When the sampling frequency is below the nyquist rate, as we see in the bottom half of \autoref{fig::explanation}, \emph{aliasing} occurs which distorts the PSD and prevents recovering the original signal. \autoref{fig:background_nyquist} demonstrates these effects on an example signal.

The implication of this theorem in practice is that when a monitored signal is sampled at or above the Nyquist rate of that signal, then operators can rest assured no information is being lost due to sampling. On the other hand, when aliasing occurs, the extent of the information loss depends on the difference between the PSD of the aliased signal and that of the original. The impact of information loss depends on the application using the measurements and operators must determine what level of aliasing (if any) is acceptable.





\begin{figure*}[!t]
    \centering
    \begin{subfigure}[b]{0.16\textwidth}
        \centering
        \includegraphics[width=\linewidth]{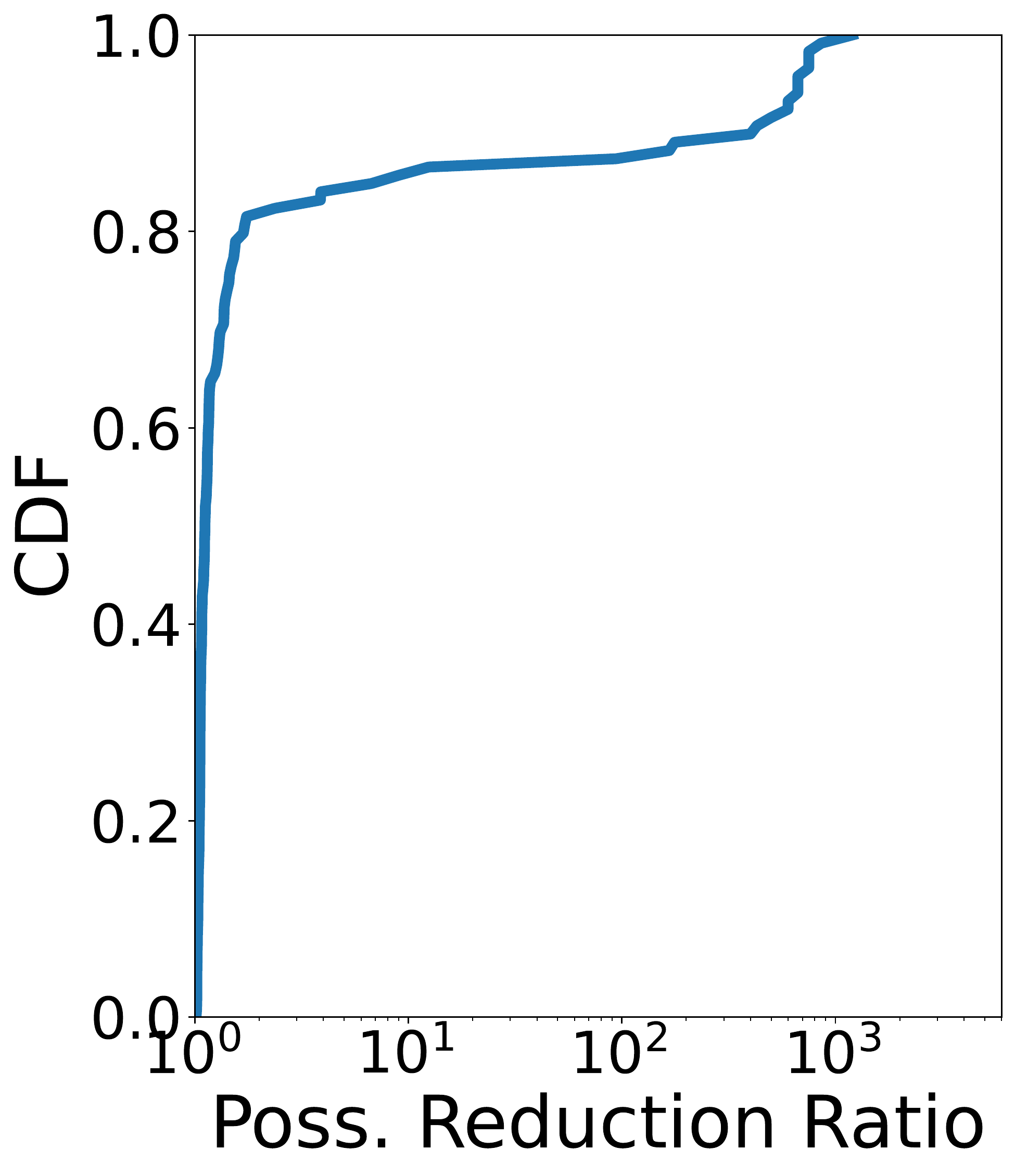}
        \caption{5-pct CPU util}
    \end{subfigure}
    \begin{subfigure}[b]{0.16\textwidth}
        \centering
        \includegraphics[width=\linewidth]{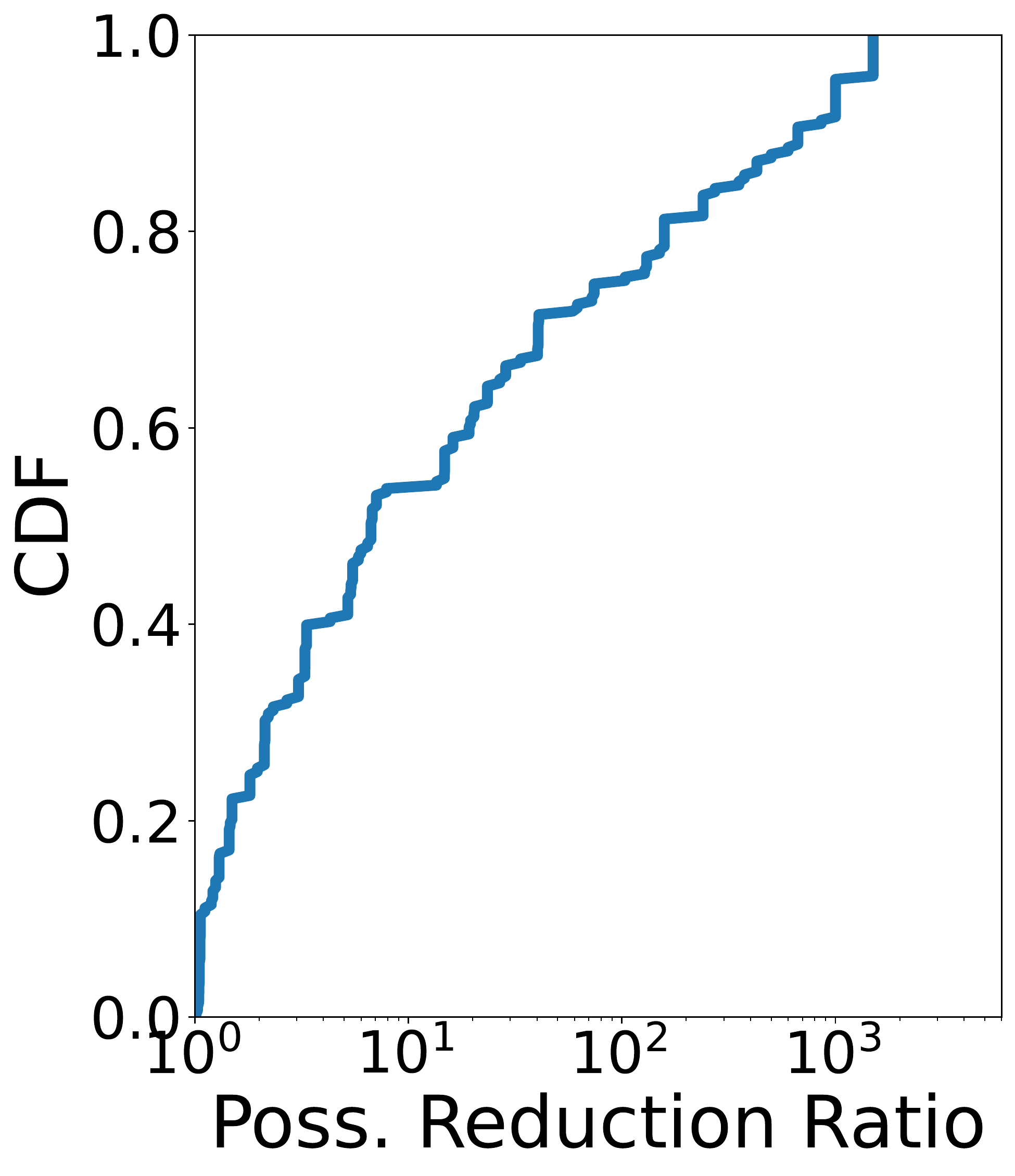}
        \caption{FCS errors}
    \end{subfigure}
    \begin{subfigure}[b]{0.16\textwidth}
        \centering
        \includegraphics[width=\linewidth]{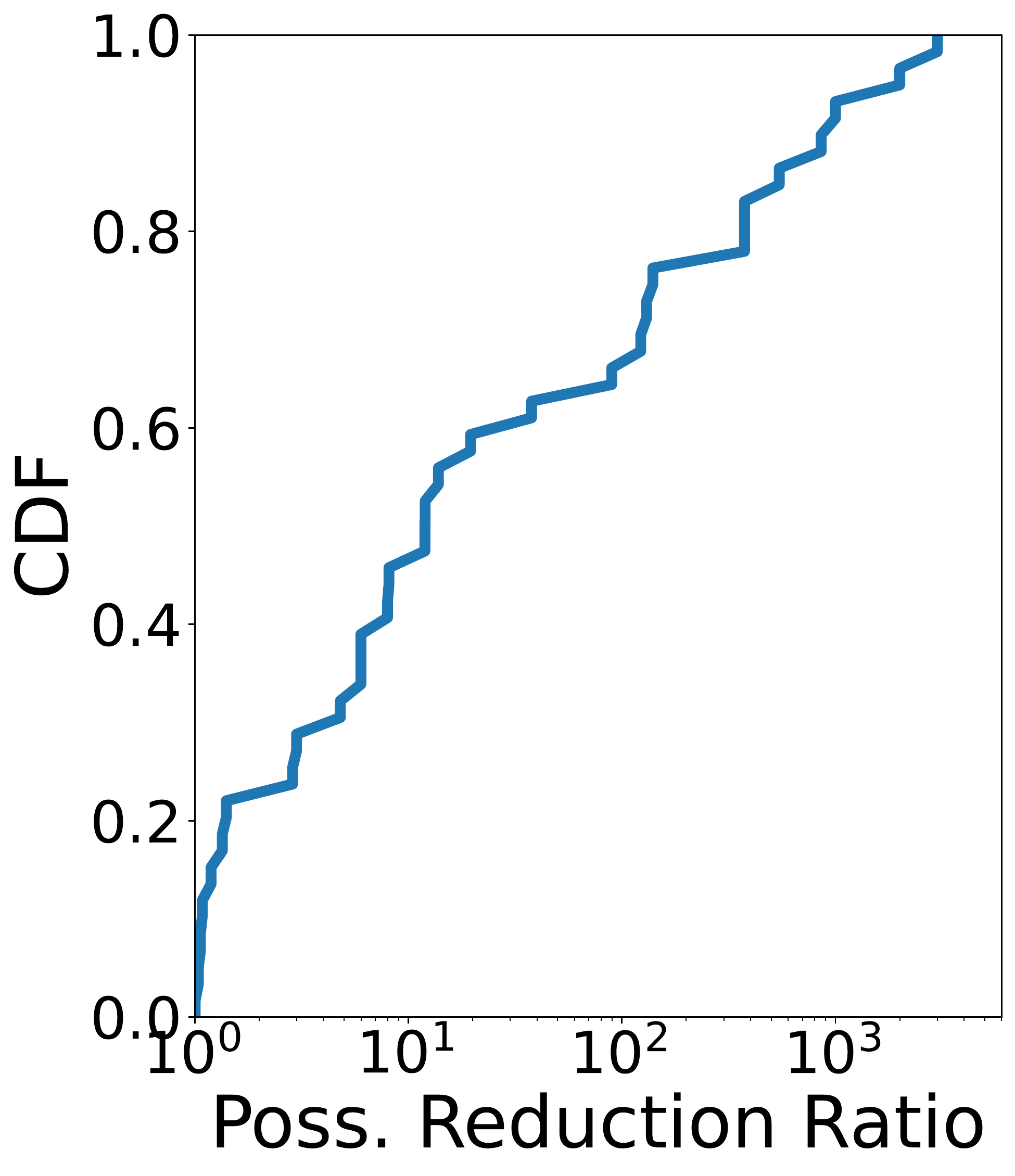}
        \caption{In-bound dis}
    \end{subfigure}
    \begin{subfigure}[b]{0.16\textwidth}
        \centering
        \includegraphics[width=\linewidth]{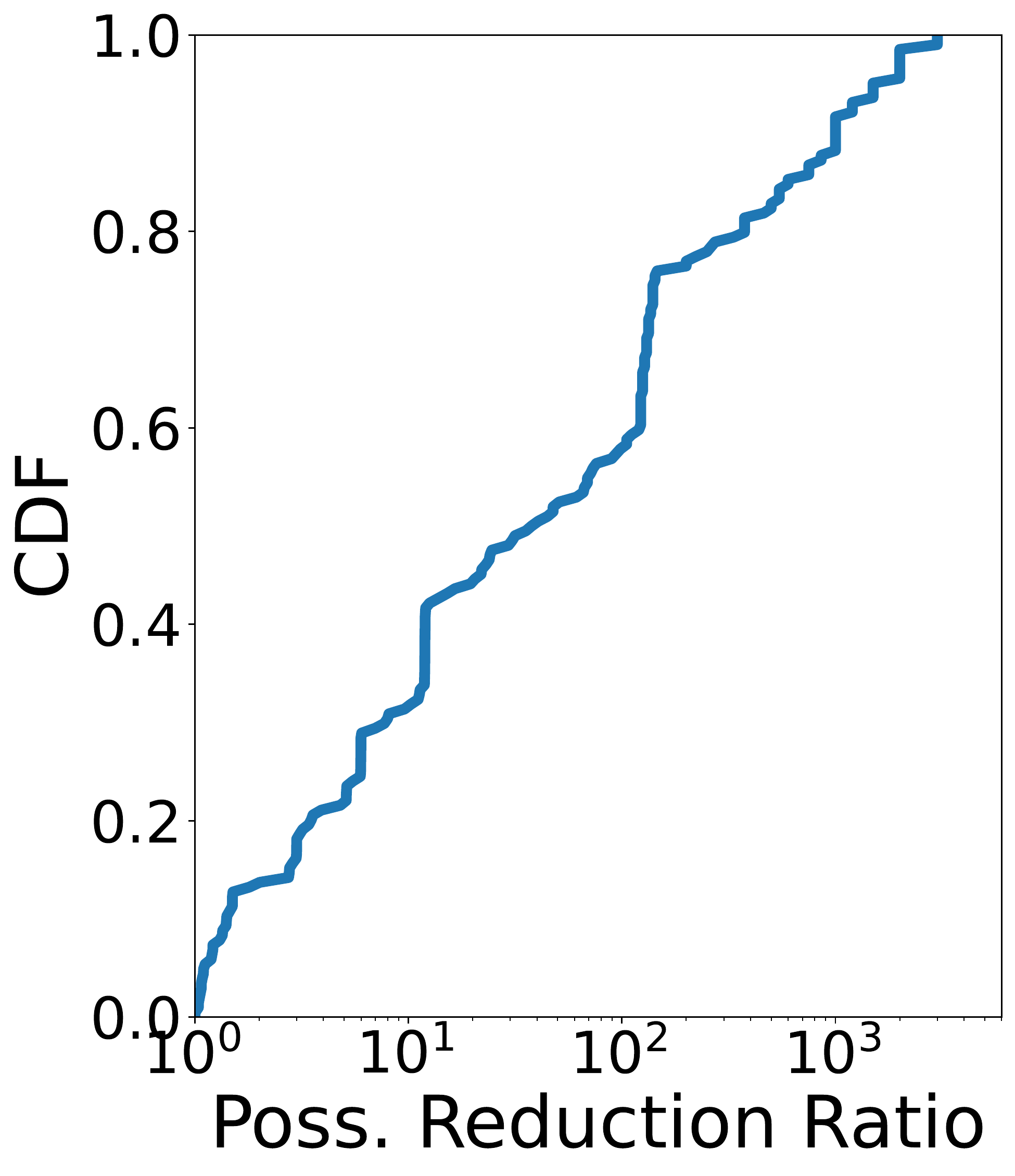}
        \caption{Link util}
    \end{subfigure}
    \begin{subfigure}[b]{0.16\textwidth}
        \centering
        \includegraphics[width=\linewidth]{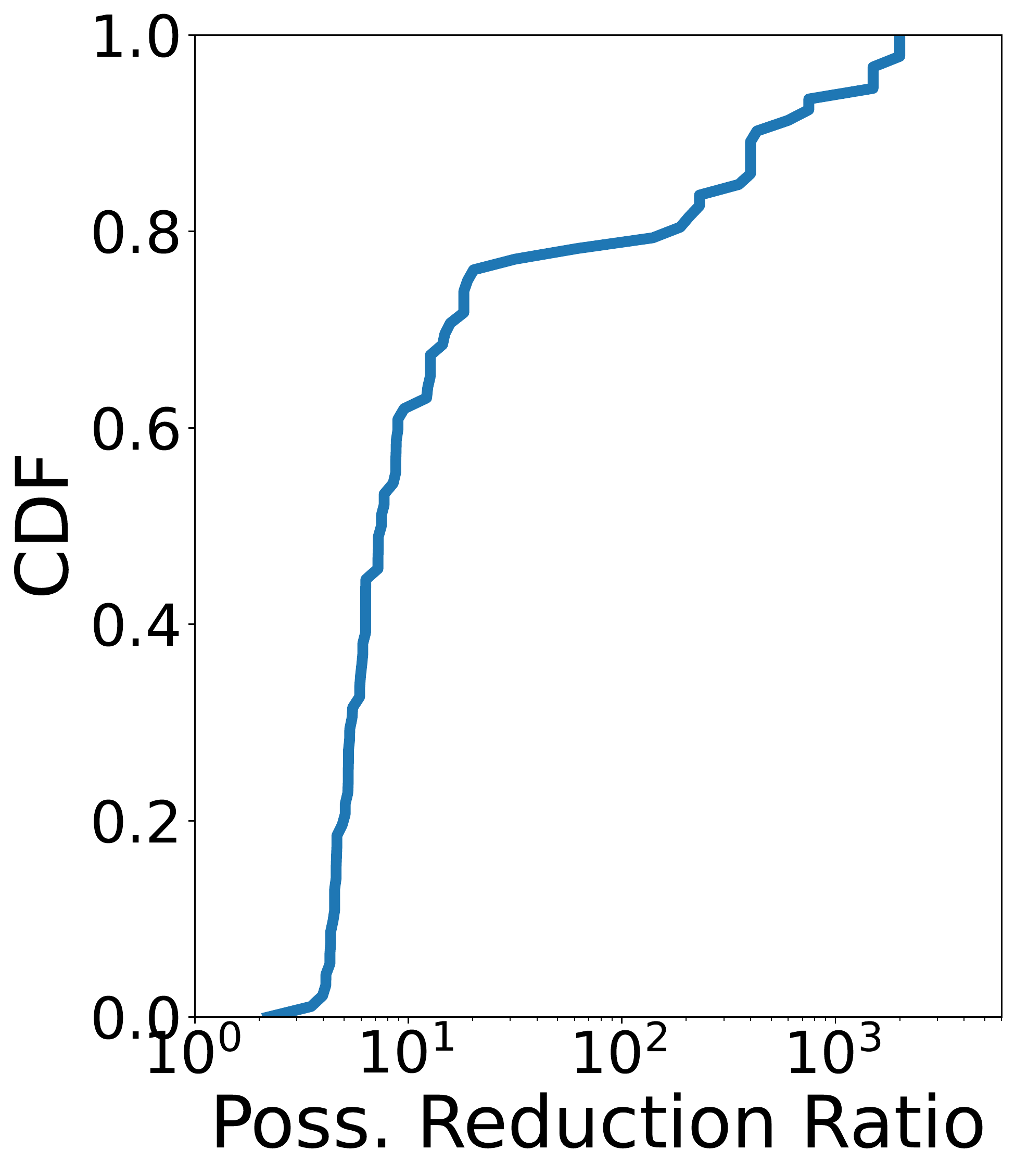}
        \caption{Lossy paths}
    \end{subfigure}
    \begin{subfigure}[b]{0.16\textwidth}
        \centering
        \includegraphics[width=\linewidth]{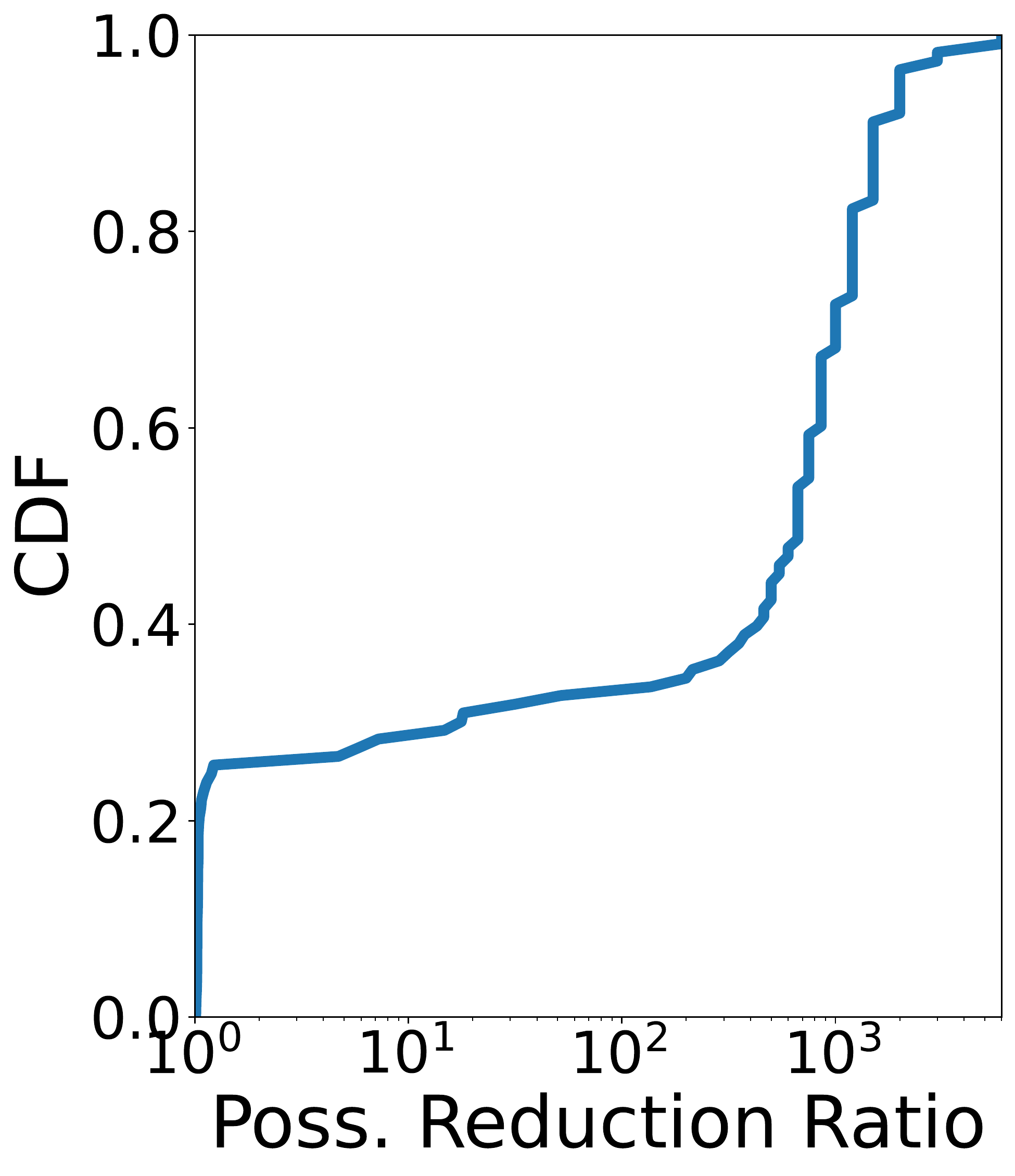}
        \caption{Memory usage}
    \end{subfigure}

    \begin{subfigure}[b]{0.16\textwidth}
        \centering
        \includegraphics[width=\linewidth]{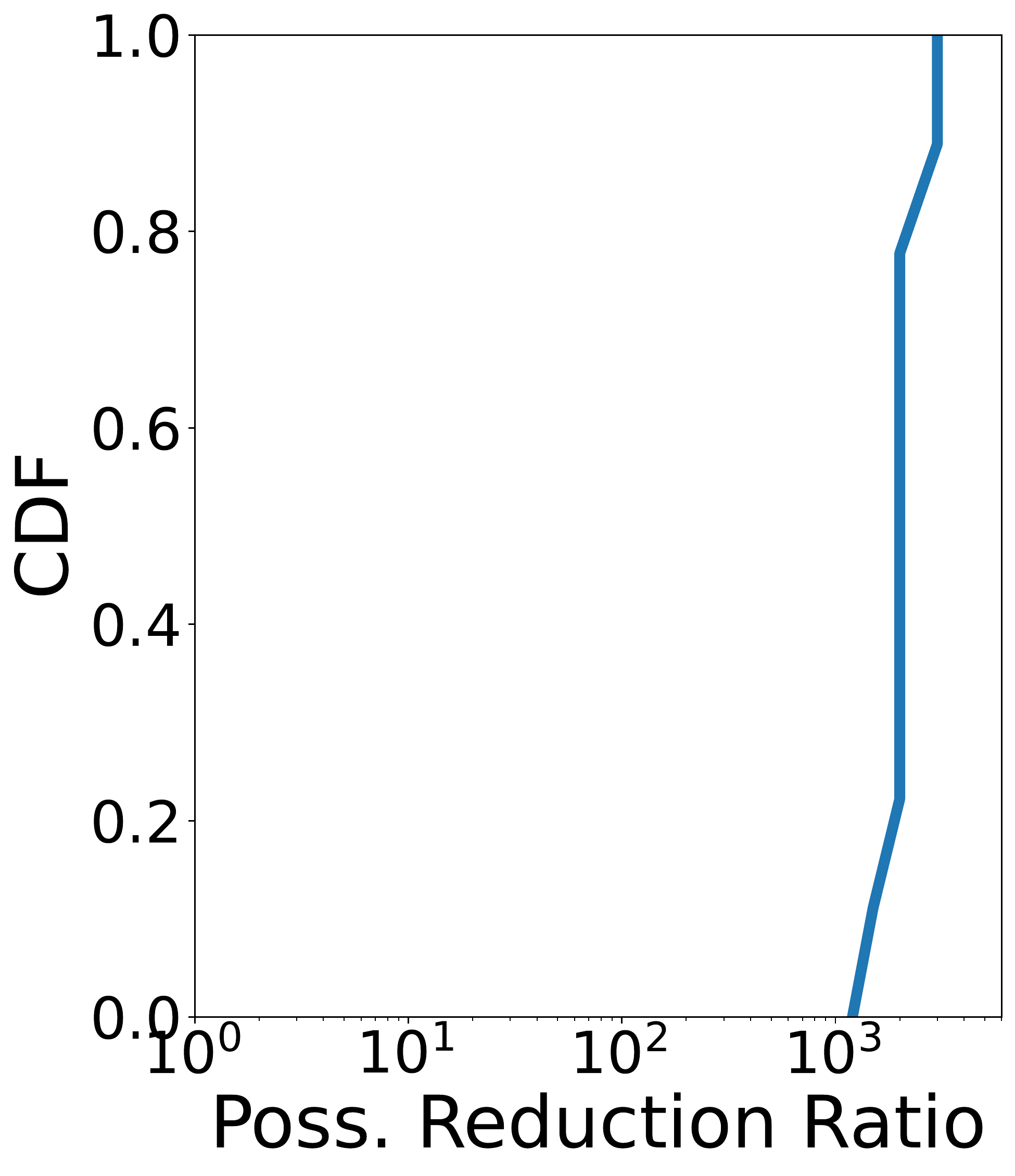}
        \caption{Multicast bytes}
    \end{subfigure}
    \begin{subfigure}[b]{0.16\textwidth}
        \centering
        \includegraphics[width=\linewidth]{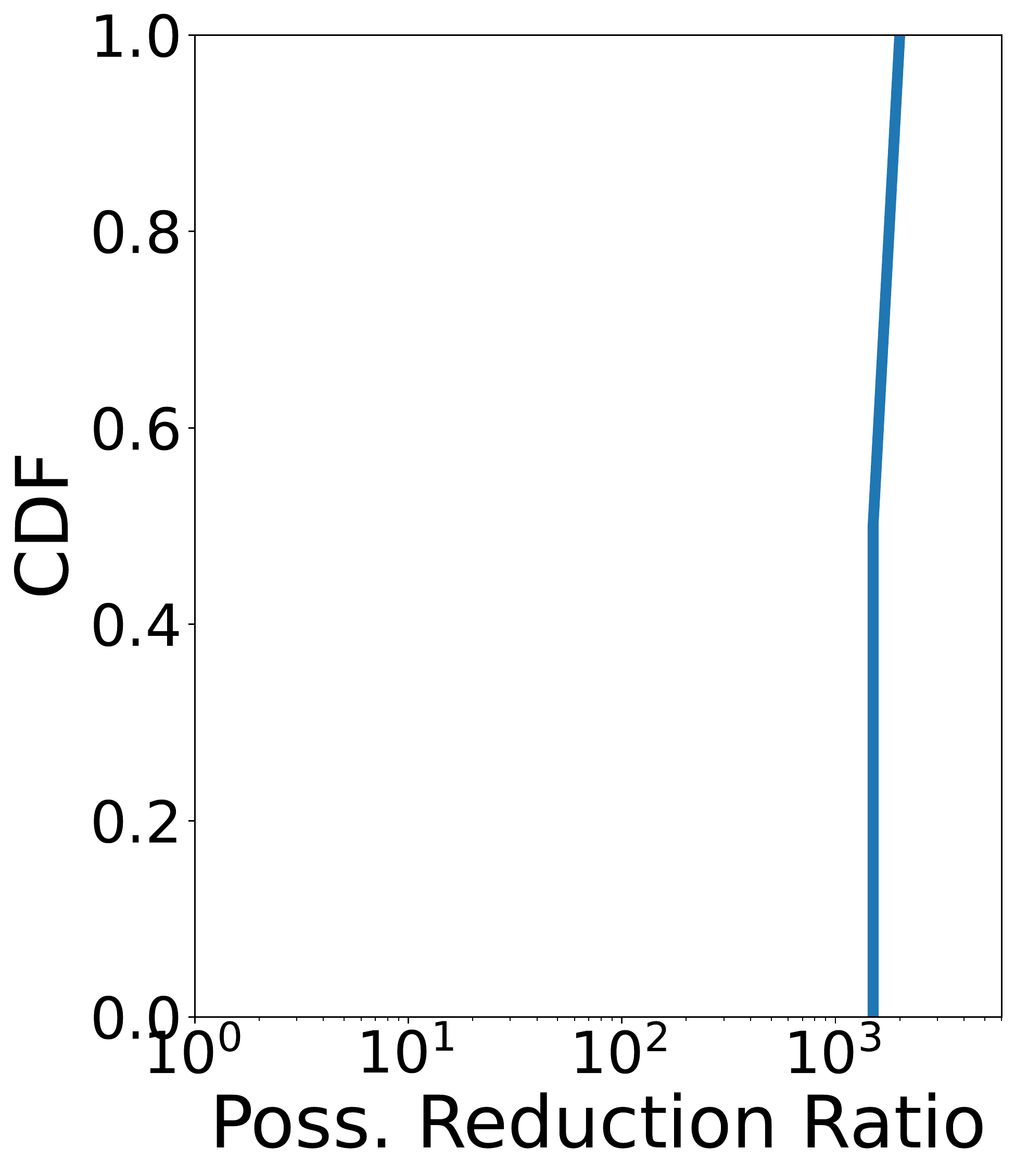}
        \caption{Multicast drops}
    \end{subfigure}
    \begin{subfigure}[b]{0.16\textwidth}
        \centering
        \includegraphics[width=\linewidth]{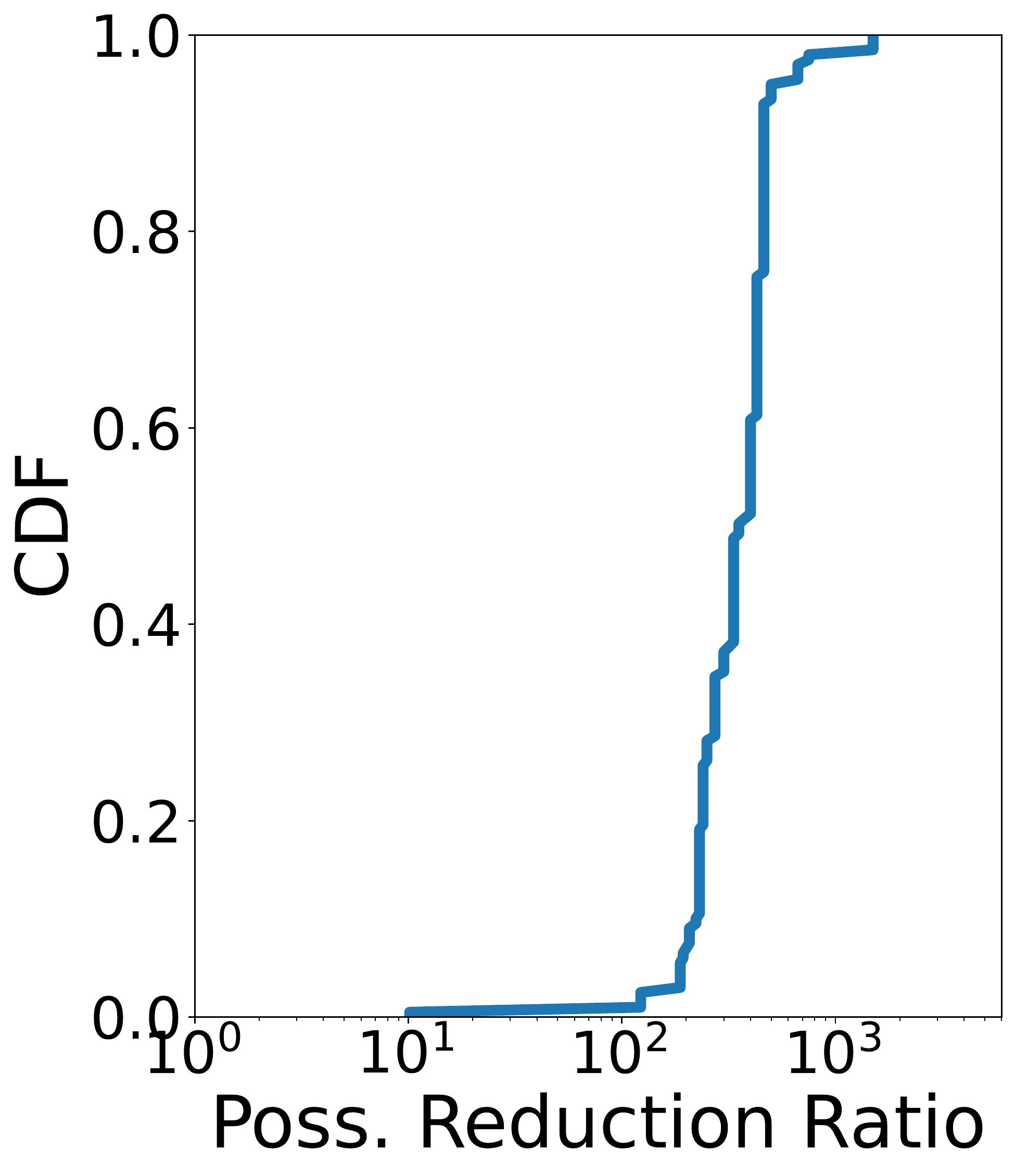}
        \caption{Peak egress BW}
    \end{subfigure}
    \begin{subfigure}[b]{0.16\textwidth}
        \centering
        \includegraphics[width=\linewidth]{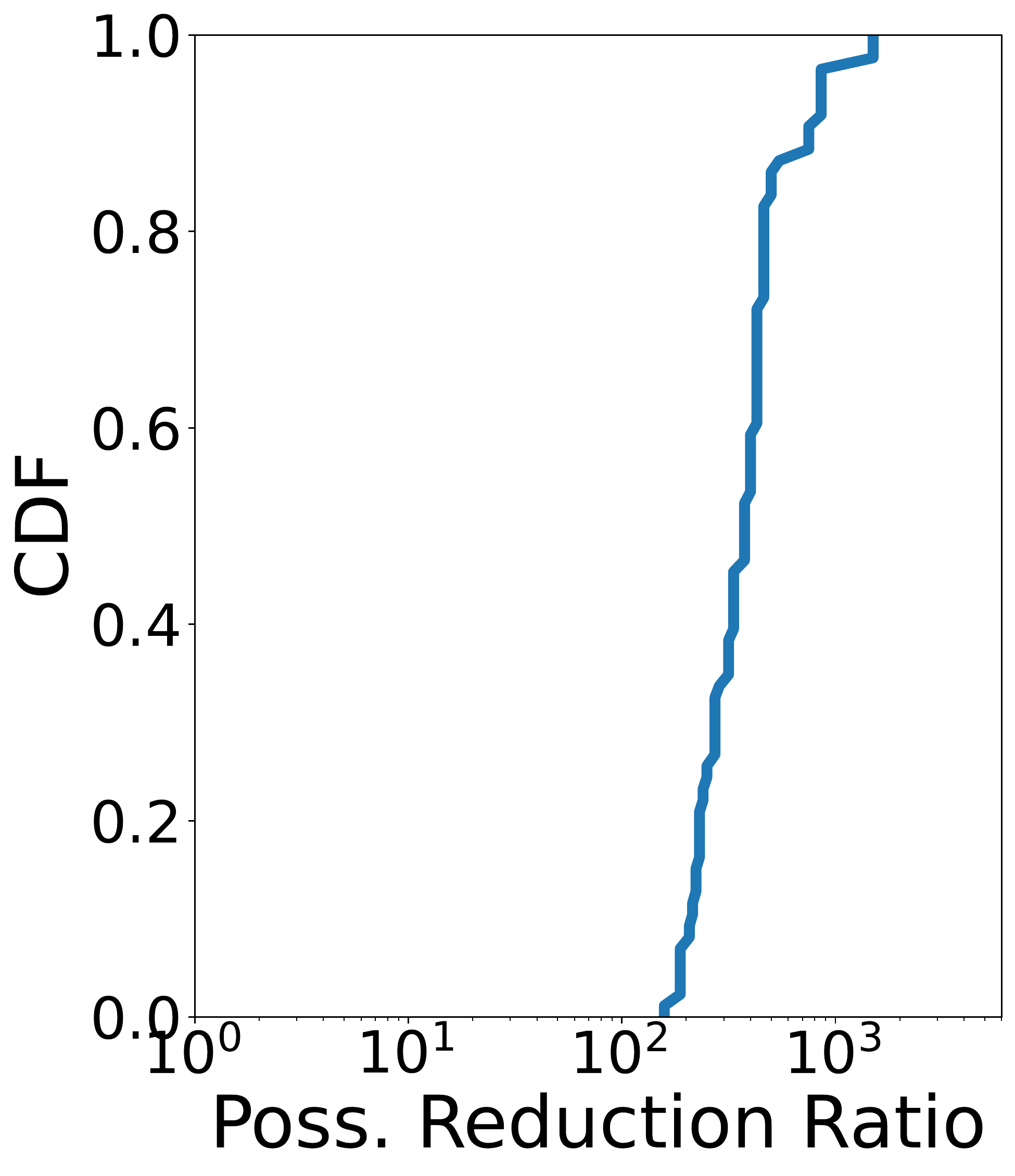}
        \caption{Peak ingress BW}
    \end{subfigure}
    \begin{subfigure}[b]{0.16\textwidth}
        \centering
        \includegraphics[width=\linewidth]{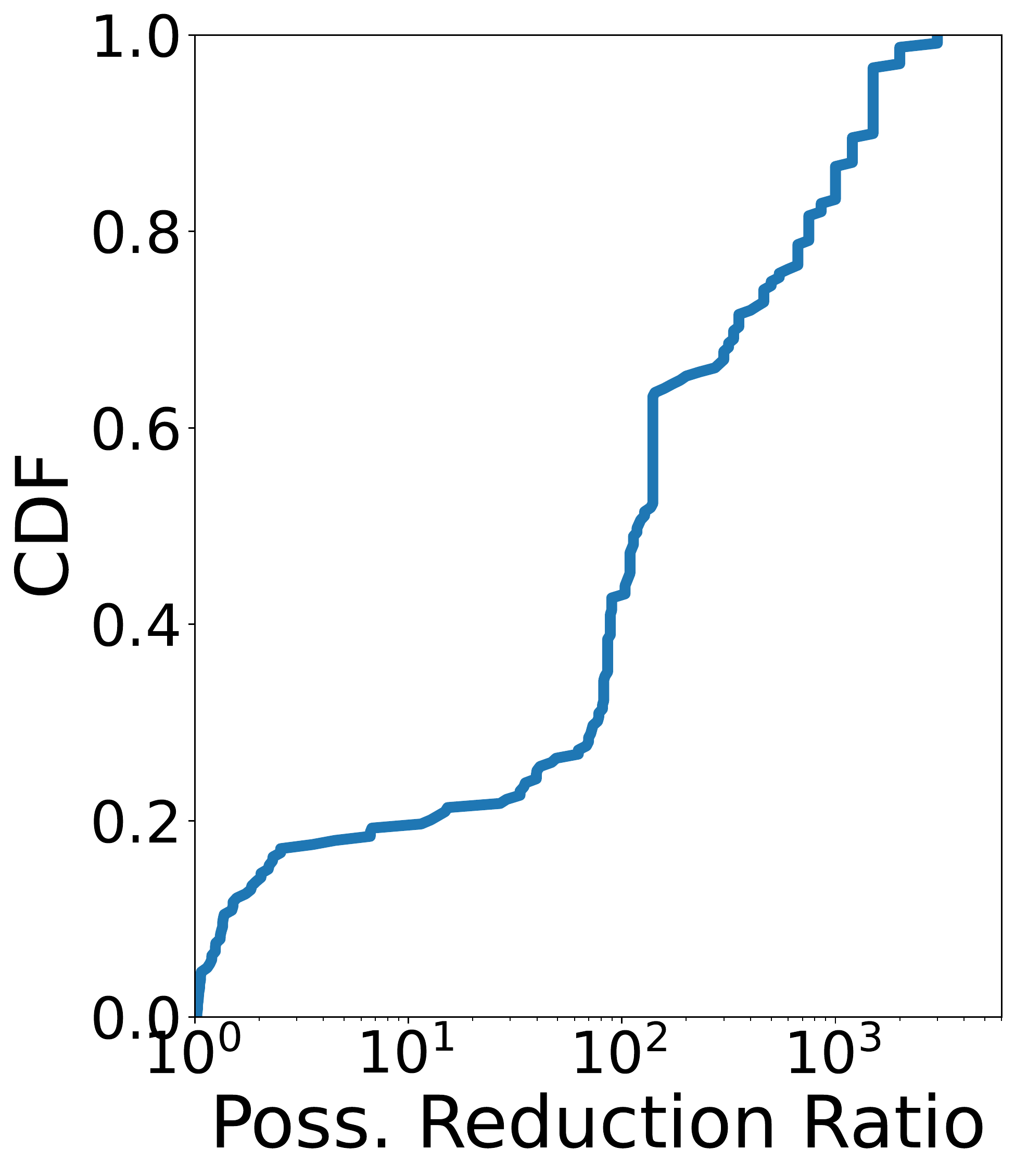}
        \caption{Temperature}
    \end{subfigure}
    \begin{subfigure}[b]{0.16\textwidth}
        \centering
        \includegraphics[width=\linewidth]{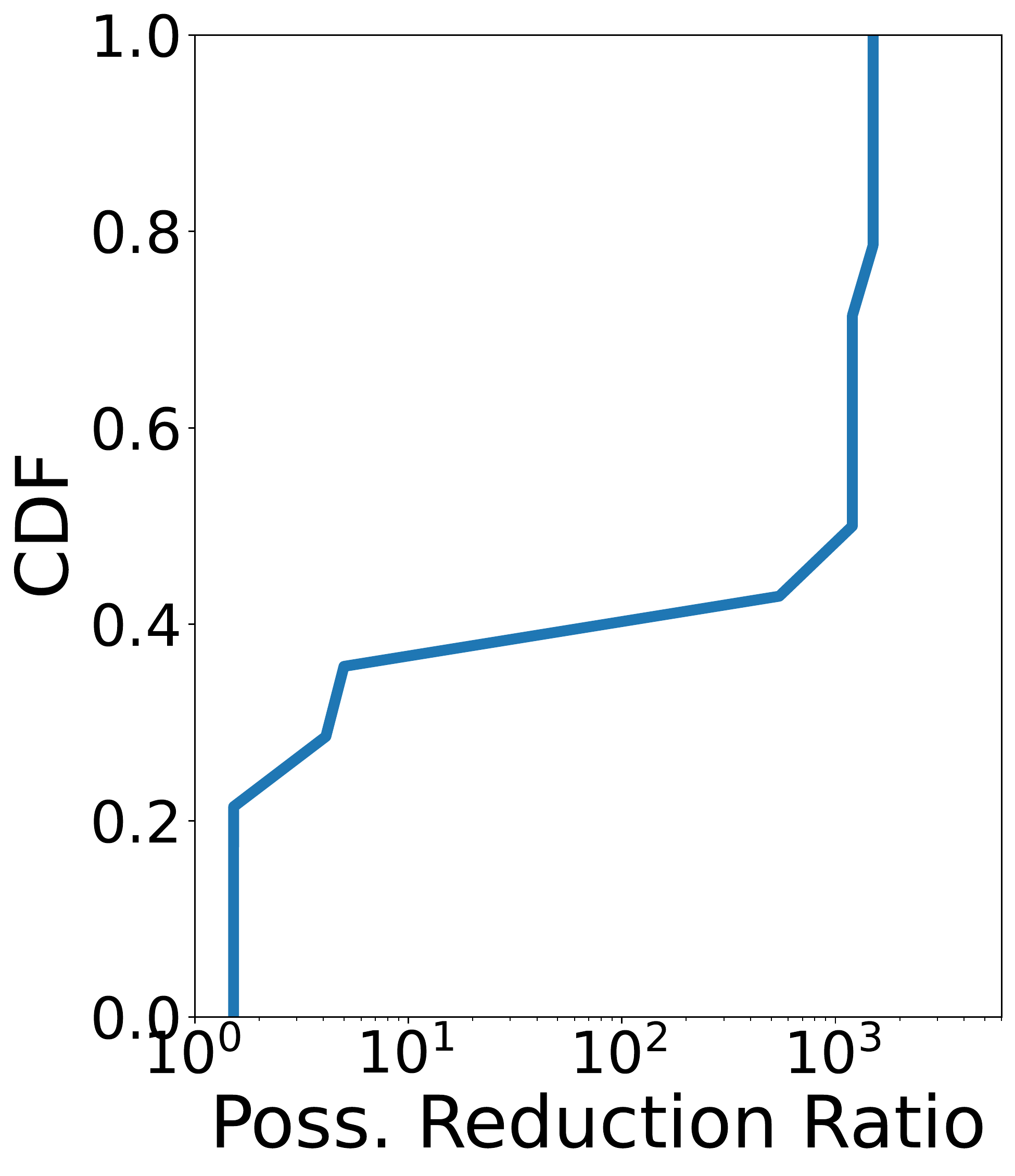}
        \caption{Unicast bytes}
    \end{subfigure}\vspace{-0.1in}
    \caption{CDFs of the ratio between the actual sampling rate and the computed Nyquist rate. Note x axes is in log scale and $x=10$ indicates $10\times$ over-sampling.  Each datapoint is one day's worth of data from a distinct device. We do not show the cases where we cannot reliably detect the Nyquist rate. \label{fig::per_system_result}}
   
\end{figure*}

\section{Quantifying the Opportunity}
\label{sec:opportunity}

Through this new lens of viewing existing network monitoring systems as collecting (sensing) signals, we can optimize many existing monitoring systems. While we may need to address many challenges before realizing these techniques in practice (see \cref{sec:practicalconcerns}), as a first step, we answer the question: what do we stand to gain if we use them?

To answer this question, we examine the network of a major cloud provider and the monitoring systems it has deployed. We describe why many of the existing monitoring systems this provider (and others similar to it) have deployed are sub-optimal.
For this discussion, we will restrict our focus to monitoring systems that periodically poll a numeric metric.
Later, we introduce opportunities for taking these observations further and applying them to other types of monitoring systems our community has developed.


\subsection{Today's Monitoring Systems}

Every aspect of the task of monitoring---collection, transmission, analysis, and storage---all consume resources that, when considering the scale of modern data centers, represent a non-negligible overhead. In an effort to reduce the costs of monitoring, for non-event-based data, today's systems generally sample their target metrics periodically/randomly so as to capture the gist of the metric without sacrificing too many resources.
Examples include systems that periodically poll switch counters~\cite{snmp,ubursts,speedlight}, sample packets to construct flow records~\cite{sflow,netflow}, or send packets through the network to extract the instantaneous latency of the network~\cite{pingmesh,INT}.

The degree of sampling for the majority of these systems is entirely arbitrary~\cite{ubursts,privateeye}.
Often, the sampling rates are not governed by signal processing principles but chosen based on defaults and vague `gut feelings' about the desired granularity of the data and the system and network-level overheads (e.g., in PrivateEye~\cite{privateeye} this is the cost of locking the virtual switch's flow-table and the volume of data that would have to be aggregated at the central collector, or in~\cite{pingmesh} it's the network overhead of sending too many pings). For such monitoring systems aliasing effects are never evaluated and the chosen granularities are never re-considered.

The end result is that most of these systems are either (a) over-sampling (increasing overheads and wasting resources) or (b) under-sampling without any idea of how much information is lost.
This has significant implications on the resource usage of existing monitoring systems.


\subsection{Case Study: A Large Cloud Provider}
\label{sec:casestudy}

We demonstrate empirically the opportunity afforded by applying the Nyquist-Shannon theorem.
The monitoring systems we study encompass a wide range of monitoring systems in a large cloud provider including device temperature, packet drops, FCS errors~\cite{corupt}, and link utilization.

Our goal is to identify the Nyquist rate of each of the signals these systems monitor. To do so: (a) for a given trace (where each trace is the data for a metric being measured on a single device), we compute the FFT and compute the total energy in the signal---the sum of the PSD across all FFT bins; (b) we add the PSD components in each FFT bin until we reach $99\%$ of the total energy in the signal computed in (a). If we need {\em all} bins of the FFT to achieve $99\%$ of the total energy we conclude the signal is probably already aliased and record -1 as the Nyquist rate; (c) otherwise, we report twice the frequency at which we capture $99\%$ of the total energy of the signal as the Nyquist rate. 

The approach above can uncover the Nyquist rate of the underlying signal {\em{if }} the measured trace is sampled above its Nyquist rate. We find this to be the common case in our traces. The converse case is challenging because when aliasing has already happened in a measured signal~(e.g., the bottom right portion of~\cref{fig::explanation}), the method above does not reliably produce the Nyquist rate. We discuss how to detect aliasing~(under-sampling) in~\cref{sec:practicalconcerns}. Our choice of the $99\%$ cut-off on total energy is a workaround to compensate for measurement noise. Using a higher parameter value such as $99.99\%$ would increase our estimate of the Nyquist rate and reduce performance gains but, in our experience, does not necessarily lead to a lower reconstruction error since the delta that is being captured is often just the noise.

In practice, monitoring systems do not produce perfectly sampled signals---samples are not always spaced at equi-distant points in time.
In such situations, we pre-clean the signal using nearest neighbor re-sampling~\cite{thevenaz2000image,Resampling2000}; that is, we add values for missing samples based on nearby samples.


\begin{figure}[t]
\centering
\includegraphics[width=\linewidth]{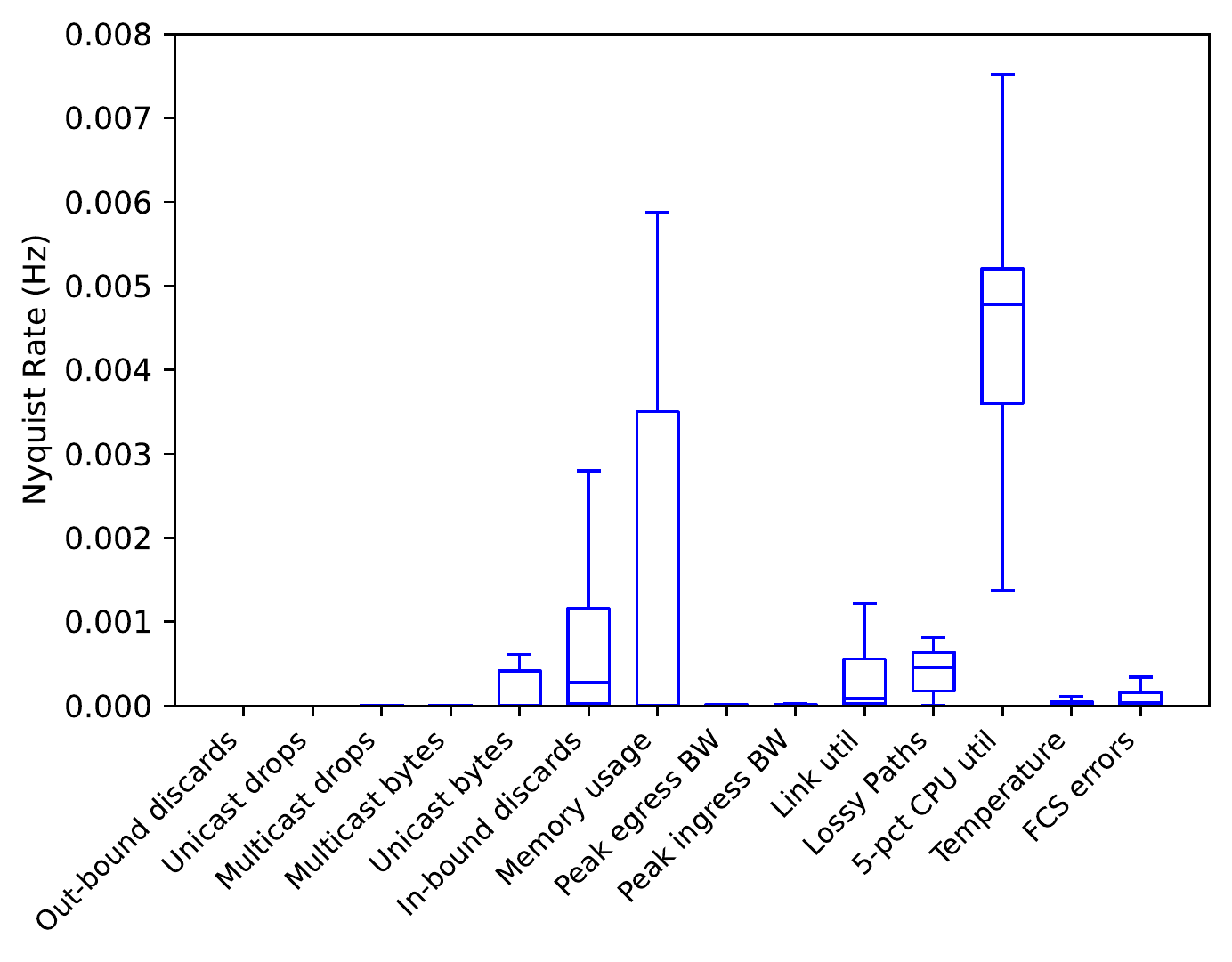}\vspace{-0.2in}
\caption{A box plot of the Nyquist rate of each monitoring system.}\vspace{-0.05in}
\label{fig::box_plot}
\end{figure}

\autoref{fig::per_system_result} reports for each measured statistic the CDF of the ratio between the current sampling and the Nyquist rate we identified through the above approach.
The ratio indicates the degree to which we are currently oversampling the underlying signal.
We observe that in $20\%$ of the examples the sampling rate can be reduced by a factor of $1000\times$. 


In total, we studied $1613$ metric and device pairs ($14$ distinct metrics). Of these, $89\%$ were sampling at higher than their Nyquist rate. Within a metric, the Nyquist rate varies widely across devices (\cref{fig::box_plot}). For example, for the temperature signal, the Nyquist rate ranges from  $7.99\times 10^{-7}$\,Hz to $0.003$\,Hz across the monitored devices. We also notice different Nyquist rate at different time periods on the same device. This indicates that the properties of the underlying metric vary over time and across devices and so  dynamically adapting the measurement rate would be useful. Furthermore, adaptation must be quick because under-sampling would lead to aliasing and information loss.  We discuss some relevant prior work~\cite{penny2003letterto} and propose a dynamic sampling method in \cref{sec:practicalconcerns}; \cref{fig::temp} shows an example of how the proposed approach would work in practice.


Finally, in our experiments the existing sampling rate is below the Nyquist rate of the underlying signal in about $11\%$ of the metric-device pairs. These cases require more careful inspection; it is possible that existing monitoring systems must increase their sampling rates in order to fully capture these metrics but we want to also rule out issues such as measurement noise, data loss or data corruption that may have lead to an incorrect assessment of the Nyquist rate. We defer doing so to our future work.




%

\section{Toward Dynamic Monitoring}
\label{sec:practicalconcerns}

In the previous section, we showed that there is a significant opportunity to save resources in deployed monitoring solutions by analyzing the signals produced by those systems.
In some cases, the actual measurement may be inexpensive relative to the cost to store the metric or the cost of downstream analysis; in such cases, we can use the above techniques a posteriori, i.e., measure  at a high rate, compute the nyquist rate over the measurements and store or present for later analysis only the measurements that are re-sampled at the lower nyquist rate. More generally, acquiring the measurements may itself be a large expense and we discuss a new dynamic sampling method below to address this case.
%
The dynamic sampling method also applies when the Nyquist rate of a metric varies across devices or across time~(recall the cases that we saw in \cref{sec:opportunity}); we do not fully understand the reasons for these changes but nevertheless use dynamic sampling to ensure robustness to changes to the nyquist rate of the underlying signal.


A complete solution to these challenges necessitates further exploration; however, the signal processing field again provides some initial avenues of exploration.
We outline potential components to a solution based on those principles.

\subsection{Detecting aliasing}
\label{ss:detect}
One critical component to a solution is a method to detect when a chosen sampling rate has dipped below the Nyquist rate of the underlying signal.
For this, prior work has previously proposed possible solutions, e.g., Penny et al.~\cite{PENNY2003473}.
In that work, the authors propose to sample at two distinct frequencies, $f_1$ and $f_2$, where $f_1 > f_2$ and $\frac{f_1}{f_2}$ is not an integer.
If aliasing occurs, i.e., the underlying signal has frequency terms that are larger than $\frac{f_2}{2}$, then comparing the discrete fourier transforms of the two sampled signals would show discrepancies; for example, frequencies below $\frac{f_2}{2}$ will match in both spectra but the higher frequencies will not match.\footnote{this requires $f_2$ to not be a factor of $f_1$ as conditioned above.} A complicating aspect here is noise at higher frequencies, but noise especially of a small amplitude can be filtered using standard techniques~\cite{TODO}.

Collecting samples at two frequencies roughly doubles measurement cost but we still expect sizable net benefit since, as we saw in~\cref{sec:opportunity} current systems, over-sample by well over $2\times$. Furthermore, after checking for aliasing, we can discard excess measurements by resampling at the identified nyquist rate.
We believe that further improvements are possible for example by using an aliasing detector that is specific to the actual frequencies and changes that appear in datacenter measurements.

\subsection{Adapting the sampling rate}
\label{ss:adapt}
Upon detecting aliasing, we increase the sampling rate. There are several possible approaches to manage the rate adaptation process and the choice among these depends on the properties of the signal (e.g., whether the high-frequency changes are one-off occurrences or sustained shifts) and the requirements for the measurement (e.g., how important is it to capture every spike in Nyquist frequency and how costly is it to oversample).

Consider the problem of quantifying link failures by sampling frame checksum~\cite{corupt} errors.
Initially, we do not know the Nyquist rate of the underlying signal and so we must probe, i.e., multiplicatively increase the measurement rate along witht he method in~\cref{ss:detect}, i.e., measuring at two different frequencies to detect aliasing. While aliasing persists, we remain in probe mode. Once we no longer detect aliasing, we use the method in in~\cref{sec:casestudy} which will successfully identify the Nyquist rate of the signal.


We can add memory into the system and/or leverage temporal stability to make adaptation faster and less expensive. If the frequency increases exhibit temporal locality (as it does in, e.g., fail-stop or link-flap scenarios), we can optimize the system by also adaptively decreasing the sampling rate if we observe the Nyquist rate returning to a lower value.
We can even `remember' previous maximum Nyquist rates to ramp up more quickly in the future.
Similarly, we may be able to learn information about applications' Nyquist shift distributions from other (oversampled) datasets from the same application.
Again, the optimal strategy will vary depending on the properties of the signal.

Perhaps the most challenging scenario for such a system is how the system should handle a first-of-its-kind event.
Maintaining ample headroom may be helpful in these cases (many of the deployed systems we examine in \cref{sec:opportunity} are already sampling at rates well above the Nyquist rate).
Additional mitigations are a subject of future work.
We note, however, the most critical issues in modern systems are those that recur---after all, ongoing fires are typically a higher priority for operators to debug than events that occur only once and never again.
An adaptive strategy will eventually be able to detect this more severe category of issues.

\subsection{Quantization Noise and Reconstruction}
\label{ss:quantization}
\begin{figure}
\centering
\includegraphics[width=\linewidth]{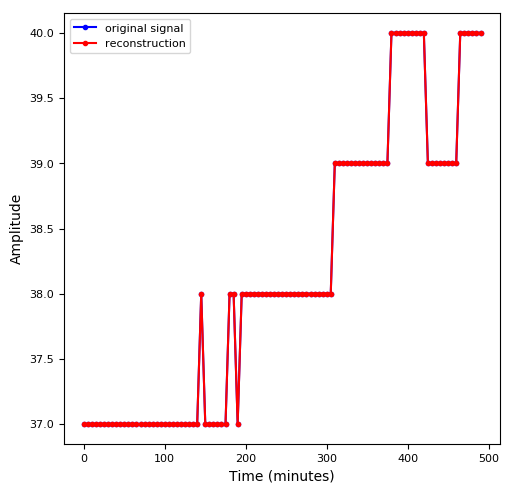}\vspace{-0.2in}
\caption{Comparing an actual temperature signal in blue~(sampled every $5$ minutes) with the signal in red that was downsampled to the nyquist rate and then upsampled back again just for the purpose of comparison. The L2 distance between these signals is $0$. Here, we used the method in~\cref{ss:adapt} to dynamically adapt the sampling rate; the inferred nyquist rates are shown as well in~\cref{fig::temp_nyquist}.\label{fig::temp}}\vspace{-0.1in}
\end{figure}

In practice, measurement readings are quantized. For example, a temperature sensor may emit readings that are rounded to the nearest integer. Such quantization adds noise which in the frequency domain appears at higher frequencies; the larger the quanta relative to the range of values that a signal can take, the higher the noise level. Quantization noise impacts our techniques in a few different ways: (a) identifying the true Nyquist rate of a signal becomes more challenging; and (b) upsampling and recovering the signal after it is downsampled.
For (a) we use the thresholding approach proposed in \cref{sec:casestudy} so as to discard higher-order frequencies introduced by quantization.
For (b) we can add the same quantization in order to recover the signal more accurately. However, in such cases the signal is no longer ``perfectly recoverable'' and the recovered signal may be slightly different from the original. \cref{fig::temp} shows the effectiveness of this approach in the context of the temperature signal. 

Using the Nyquist principle and the adaptive sampling approach described above, we are able to reduce the overhead on the monitoring system. However, in order to reconstruct the signal, operators would have to pass the signal through a low-pass filter (for example, by taking an FFT of the sampled signal, setting all frequency components above $f_0$ to $0$ and then taking the IFFT). This reconstruction takes time and may not be acceptable to applications that expect low-latency. However, in many cases this reconstruction cost is acceptable. For instance, in machine learning models where some delay in recovering the data in return for higher-fidelity is desired.

\begin{figure}
\centering
\includegraphics[width=\linewidth]{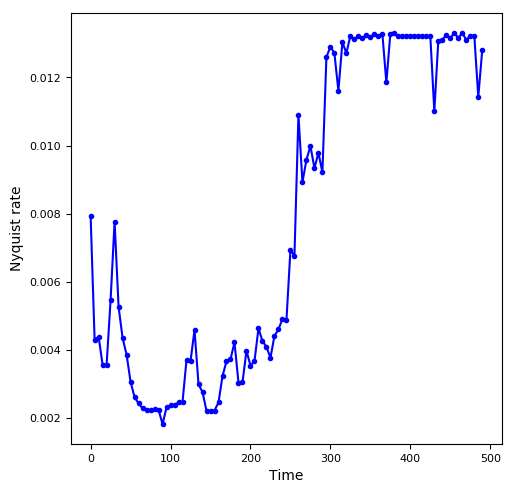}\vspace{-0.2in}
\caption{The inferred Nyquist rates over time for the signal depicted in~\cref{fig::temp}. The timestamps mark the beginning of the moving window. We use a step of 5 minutes for the moving window and a window size of 6 hours.\label{fig::temp_nyquist}}\vspace{-0.1in}
\end{figure}
\section{Related Work}

In this paper, we proposed a different perspective on how we approach network monitoring.
The research we propose builds on top of the vast array of knowledge in the field of signal processing and relates to the following categories of work in signal processing and network monitoring:

\heading{Signal processing theory and its applications to networking} The Nyquist--Shannon sampling theorem is ubiquitous and has been applied in several other domains~\cite{farrow2011nyquist,romo2018nyquist,zamaruiev2017use} and been expanded in its applicability~\cite{song2012improved}.
Prior work has also focused on how to detect aliasing in practice~\cite{penny2003letterto}.
We build on all of this work.


In network monitoring, the most closely related work are the existing, but isolated examples of signal processing techniques in systems and networks.
These include techniques like compressive sensing, which shares both motivation and intuition with our work. 
In \cite{265057}, for example, authors regard flow statistics as signals to design new sketch algorithms that bound the information loss of the system.
Similarly, in~\cite{zhang2005network} the authors use FFTs to identify network faults. Others, characterize the underlying properties of network traffic efficiently and accurately~\cite{642356,10.1007/978-3-540-25969-5_9,10.1007/3-540-45749-6_33,benson2010network,STOEV2005423}. Stoev et al.~\cite{STOEV2005423} developed analyses to better characterize the properties of Internet traffic in the presence of high-frequency oscillations and non-stationarities.
Finally,~\cite{10.1007/3-540-45749-6_33} presents an algorithm that can track categories of traffic with frequencies above a certain value with high probability.

These examples are complementary to our work, and they serve as examples of the potential of viewing the data center as a collection of digital signals.


\heading{Network monitoring systems} More generally, many others have also tried to address the problem of scalability/over\-head in both the collection and storage of network measurement.
Sketches like the one cited above~\cite{265057} and~\cite{ivkin2019qpipe,moshref2015scream} are one approach that researchers have used to do so. Sketches reduce the space required for storing data plane statistics, but those statistics must still be sampled when summarized for users or stored for later analysis.

There is also a vast body of various network monitoring systems, from simple, periodic ones like~\cite{pingmesh,corupt}, to those that are further optimized~\cite{moshref2014dream,tammana2016simplifying,holterbach2019blink,moshref2016trumpet}. It may be possible to improve the efficiency of these systems as well by re-framing the monitored information as time-varying signal and bringing to bear the vast theory of signal processing as already demonstrated by~\cite{265057}.

Finally, many network protocols also measure (most prominently TCP) and continuously monitor the network. Here too we may be able to improve efficiency through our new lens of viewing the information they monitor (in the case of TCP this would be the network round trip time, and the bottleneck bandwidth) as a continuous or discrete-time signals.

In summary, this paper advocates for a broader re-examin\-ation of modern data center systems with an eye toward directly improving the efficiency of those systems by bringing to bear the vast theory present in signal processing. More specifically, we pose the significant gains possible by bringing to bear the rather simply stated (but more nuanced in practice) Nyquist-Shannon sampling theorem.

\section{Conclusion and Future Work}

Monitoring systems are a crucial part of ensuring high availability for data centers.
Yet, efficient measurements based on information theoretic concepts are not widely studied. We argue that the next stage of network monitoring should go beyond vague ‘gut feelings’ about the granularity of data and instead leverage concepts from signal processing to help avoid wasteful collection.

We show how Nyquist-Shannon might provide guidance to optimize existing monitoring systems and propose a straw-man approach for adaptive sampling.





\heading{Beyond numbers}
In our experiments and discussions above we focused on monitoring systems that already fit nicely into the Nyquist model of sampling: they periodically sample a numeric metric at a fixed frequency.
Further research is needed to apply the same techniques to other more complex systems such as~\cite{moshref2014dream,moshref2015scream,moshref2016trumpet,265057,tpprof}, which may measure events, sets of metrics, or text representations.
We defer solving this problem to future work.

\heading{Multivariate signals}
Many applications may monitor and use multiple different signals.
The correlation and joint distribution of these signals may be important to such applications.
As long as we sample each individual signal at a rate higher than its Nyquist rate, we can recover the original signal and preserve any correlations.
However, if the Nyquist rate changes frequently and the system attempts to use the dynamic increase/decrease approach of \cref{sec:practicalconcerns}, the algorithm we presented may encounter pathological cases.
For example, even the sampling of every individual signal converges to the Nyquist rate, the global optimal is not guaranteed.

Luckily, the signal processing community has worked on extending the Nyquist theorem to multivariate signals~\cite{nathan1978plain, freeden2018gaussian}. Incorporating these extensions into a practical system is left to future work.

\heading{Beyond Nyquist}
%
Although we focused on the Nyquist--Shannon theorem in this paper and how it can help improve the efficiency of deployed monitoring systems, it is only one example of a technique from the realm of digital signal processing that can be applied to data centers.

Another is ergodicity.
In statistical mechanics, an ergotic process is one in which the statistical properties of a sufficiently long random sample of the process are equivalent to the properties of a random sample of a statistical ensemble of the process.
As an example, consider a system that monitors the CPU utilization of all servers in a data center.
Samples from the system are ergotic if the statistical properties of a set of samples derived from a single CPU over a sufficiently long sequence of time are equivalent to those of a set of samples derived from measuring the entire fleet at once.

Operators often assume ergodicity implicitly.
One example is the practice of canarying, where an update is rolled out to a handful of servers/racks/switches to evaluate its effects before deploying the update more broadly.
Extrapolating canary results to other devices relies on ergodicity.
Does this assumption hold in practice?
How long of an observation period is required for the assumption to hold true?
Is there a way to leverage ergodicity to reduce the number of devices that we need to sample or the time for which we need to sample them?
The literature here is equally rich and should be explored.










\bibliographystyle{ACM-Reference-Format.bst} 
\bibliography{refs}


\begin{thebibliography}{34}


\ifx \showCODEN    \undefined \def \showCODEN     #1{\unskip}     \fi
\ifx \showDOI      \undefined \def \showDOI       #1{#1}\fi
\ifx \showISBNx    \undefined \def \showISBNx     #1{\unskip}     \fi
\ifx \showISBNxiii \undefined \def \showISBNxiii  #1{\unskip}     \fi
\ifx \showISSN     \undefined \def \showISSN      #1{\unskip}     \fi
\ifx \showLCCN     \undefined \def \showLCCN      #1{\unskip}     \fi
\ifx \shownote     \undefined \def \shownote      #1{#1}          \fi
\ifx \showarticletitle \undefined \def \showarticletitle #1{#1}   \fi
\ifx \showURL      \undefined \def \showURL       {\relax}        \fi
\providecommand\bibfield[2]{#2}
\providecommand\bibinfo[2]{#2}
\providecommand\natexlab[1]{#1}
\providecommand\showeprint[2][]{arXiv:#2}

\bibitem[\protect\citeauthoryear{Arzani, Ciraci, Saroiu, Wolman, Stokes,
  Outhred, and Diwu}{Arzani et~al\mbox{.}}{2020}]%
        {privateeye}
\bibfield{author}{\bibinfo{person}{Behnaz Arzani}, \bibinfo{person}{Selim
  Ciraci}, \bibinfo{person}{Stefan Saroiu}, \bibinfo{person}{Alec Wolman},
  \bibinfo{person}{Jack Stokes}, \bibinfo{person}{Geoff Outhred}, {and}
  \bibinfo{person}{Lechao Diwu}.} \bibinfo{year}{2020}\natexlab{}.
\newblock \showarticletitle{PrivateEye: Scalable and Privacy-Preserving
  Compromise Detection in the Cloud}. In \bibinfo{booktitle}{\emph{17th
  $\{$USENIX$\}$ Symposium on Networked Systems Design and Implementation
  ($\{$NSDI$\}$ 20)}}. \bibinfo{pages}{797--815}.
\newblock


\bibitem[\protect\citeauthoryear{Benson, Akella, and Maltz}{Benson
  et~al\mbox{.}}{2010}]%
        {benson2010network}
\bibfield{author}{\bibinfo{person}{Theophilus Benson}, \bibinfo{person}{Aditya
  Akella}, {and} \bibinfo{person}{David~A. Maltz}.}
  \bibinfo{year}{2010}\natexlab{}.
\newblock \showarticletitle{Network Traffic Characteristics of Data Centers in
  the Wild}. In \bibinfo{booktitle}{\emph{Proceedings of the 10th ACM SIGCOMM
  Conference on Internet Measurement}} (Melbourne, Australia)
  \emph{(\bibinfo{series}{IMC '10})}. \bibinfo{publisher}{Association for
  Computing Machinery}, \bibinfo{address}{New York, NY, USA},
  \bibinfo{pages}{267–280}.
\newblock
\showISBNx{9781450304832}
\urldef\tempurl%
\url{https://doi.org/10.1145/1879141.1879175}
\showDOI{\tempurl}


\bibitem[\protect\citeauthoryear{Claise, Sadasivan, Valluri, and
  Djernaes}{Claise et~al\mbox{.}}{2004}]%
        {netflow}
\bibfield{author}{\bibinfo{person}{Benoit Claise}, \bibinfo{person}{Ganesh
  Sadasivan}, \bibinfo{person}{Vamsi Valluri}, {and} \bibinfo{person}{Martin
  Djernaes}.} \bibinfo{year}{2004}\natexlab{}.
\newblock \showarticletitle{Cisco systems netflow services export version 9}.
\newblock  (\bibinfo{year}{2004}).
\newblock


\bibitem[\protect\citeauthoryear{Demaine, L{\'o}pez-Ortiz, and Munro}{Demaine
  et~al\mbox{.}}{2002}]%
        {10.1007/3-540-45749-6_33}
\bibfield{author}{\bibinfo{person}{Erik~D. Demaine}, \bibinfo{person}{Alejandro
  L{\'o}pez-Ortiz}, {and} \bibinfo{person}{J.~Ian Munro}.}
  \bibinfo{year}{2002}\natexlab{}.
\newblock \showarticletitle{Frequency Estimation of Internet Packet Streams
  with Limited Space}. In \bibinfo{booktitle}{\emph{Algorithms --- ESA 2002}},
  \bibfield{editor}{\bibinfo{person}{Rolf M{\"o}hring} {and}
  \bibinfo{person}{Rajeev Raman}} (Eds.). \bibinfo{publisher}{Springer Berlin
  Heidelberg}, \bibinfo{address}{Berlin, Heidelberg},
  \bibinfo{pages}{348--360}.
\newblock
\showISBNx{978-3-540-45749-7}


\bibitem[\protect\citeauthoryear{Farrow, Shaw, Kim, Juh{\'a}s, and
  Billinge}{Farrow et~al\mbox{.}}{2011}]%
        {farrow2011nyquist}
\bibfield{author}{\bibinfo{person}{Christopher~L Farrow},
  \bibinfo{person}{Margaret Shaw}, \bibinfo{person}{Hyunjeong Kim},
  \bibinfo{person}{Pavol Juh{\'a}s}, {and} \bibinfo{person}{Simon~JL
  Billinge}.} \bibinfo{year}{2011}\natexlab{}.
\newblock \showarticletitle{Nyquist-Shannon sampling theorem applied to
  refinements of the atomic pair distribution function}.
\newblock \bibinfo{journal}{\emph{Physical Review B}} \bibinfo{volume}{84},
  \bibinfo{number}{13} (\bibinfo{year}{2011}), \bibinfo{pages}{134105}.
\newblock


\bibitem[\protect\citeauthoryear{Freeden and Nashed}{Freeden and
  Nashed}{2018}]%
        {freeden2018gaussian}
\bibfield{author}{\bibinfo{person}{Willi Freeden} {and}
  \bibinfo{person}{M~Zuhair Nashed}.} \bibinfo{year}{2018}\natexlab{}.
\newblock \showarticletitle{From the Gaussian Circle Problem to Multivariate
  Shannon Sampling}.
\newblock In \bibinfo{booktitle}{\emph{Frontiers in Orthogonal Polynomials and
  q-Series}}. \bibinfo{publisher}{World Scientific}, \bibinfo{pages}{213--237}.
\newblock


\bibitem[\protect\citeauthoryear{Guo, Yuan, Xiang, Dang, Huang, Maltz, Liu,
  Wang, Pang, Chen, Lin, and Kurien}{Guo et~al\mbox{.}}{2015}]%
        {pingmesh}
\bibfield{author}{\bibinfo{person}{Chuanxiong Guo}, \bibinfo{person}{Lihua
  Yuan}, \bibinfo{person}{Dong Xiang}, \bibinfo{person}{Yingnong Dang},
  \bibinfo{person}{Ray Huang}, \bibinfo{person}{Dave Maltz},
  \bibinfo{person}{Zhaoyi Liu}, \bibinfo{person}{Vin Wang},
  \bibinfo{person}{Bin Pang}, \bibinfo{person}{Hua Chen},
  \bibinfo{person}{Zhi-Wei Lin}, {and} \bibinfo{person}{Varugis Kurien}.}
  \bibinfo{year}{2015}\natexlab{}.
\newblock \showarticletitle{Pingmesh: A Large-Scale System for Data Center
  Network Latency Measurement and Analysis}. In
  \bibinfo{booktitle}{\emph{Proceedings of the 2015 ACM Conference on Special
  Interest Group on Data Communication}} (London, United Kingdom)
  \emph{(\bibinfo{series}{SIGCOMM '15})}. \bibinfo{publisher}{ACM},
  \bibinfo{address}{New York, NY, USA}, \bibinfo{pages}{139--152}.
\newblock
\showISBNx{978-1-4503-3542-3}
\urldef\tempurl%
\url{https://doi.org/10.1145/2785956.2787496}
\showDOI{\tempurl}


\bibitem[\protect\citeauthoryear{Hare}{Hare}{2011}]%
        {snmp}
\bibfield{author}{\bibinfo{person}{Chris Hare}.}
  \bibinfo{year}{2011}\natexlab{}.
\newblock \bibinfo{title}{Simple Network Management Protocol (SNMP).}
\newblock
\newblock


\bibitem[\protect\citeauthoryear{Holterbach, Molero, Apostolaki, Dainotti,
  Vissicchio, and Vanbever}{Holterbach et~al\mbox{.}}{2019}]%
        {holterbach2019blink}
\bibfield{author}{\bibinfo{person}{Thomas Holterbach},
  \bibinfo{person}{Edgar~Costa Molero}, \bibinfo{person}{Maria Apostolaki},
  \bibinfo{person}{Alberto Dainotti}, \bibinfo{person}{Stefano Vissicchio},
  {and} \bibinfo{person}{Laurent Vanbever}.} \bibinfo{year}{2019}\natexlab{}.
\newblock \showarticletitle{Blink: Fast connectivity recovery entirely in the
  data plane}. In \bibinfo{booktitle}{\emph{16th $\{$USENIX$\}$ Symposium on
  Networked Systems Design and Implementation ($\{$NSDI$\}$ 19)}}.
  \bibinfo{pages}{161--176}.
\newblock


\bibitem[\protect\citeauthoryear{Huang, Sheng, Chen, Bao, Zhang, Xu, and
  Zhang}{Huang et~al\mbox{.}}{2021}]%
        {265057}
\bibfield{author}{\bibinfo{person}{Qun Huang}, \bibinfo{person}{Siyuan Sheng},
  \bibinfo{person}{Xiang Chen}, \bibinfo{person}{Yungang Bao},
  \bibinfo{person}{Rui Zhang}, \bibinfo{person}{Yanwei Xu}, {and}
  \bibinfo{person}{Gong Zhang}.} \bibinfo{year}{2021}\natexlab{}.
\newblock \showarticletitle{Toward Nearly-Zero-Error Sketching via Compressive
  Sensing}. In \bibinfo{booktitle}{\emph{18th {USENIX} Symposium on Networked
  Systems Design and Implementation ({NSDI} 21)}}. \bibinfo{publisher}{{USENIX}
  Association}, \bibinfo{pages}{1027--1044}.
\newblock
\showISBNx{978-1-939133-21-2}
\urldef\tempurl%
\url{https://www.usenix.org/conference/nsdi21/presentation/huang}
\showURL{%
\tempurl}


\bibitem[\protect\citeauthoryear{Ivkin, Yu, Braverman, and Jin}{Ivkin
  et~al\mbox{.}}{2019}]%
        {ivkin2019qpipe}
\bibfield{author}{\bibinfo{person}{Nikita Ivkin}, \bibinfo{person}{Zhuolong
  Yu}, \bibinfo{person}{Vladimir Braverman}, {and} \bibinfo{person}{Xin Jin}.}
  \bibinfo{year}{2019}\natexlab{}.
\newblock \showarticletitle{Qpipe: Quantiles sketch fully in the data plane}.
  In \bibinfo{booktitle}{\emph{Proceedings of the 15th International Conference
  on Emerging Networking Experiments And Technologies}}.
  \bibinfo{pages}{285--291}.
\newblock


\bibitem[\protect\citeauthoryear{Moshref, Yu, Govindan, and Vahdat}{Moshref
  et~al\mbox{.}}{2014}]%
        {moshref2014dream}
\bibfield{author}{\bibinfo{person}{Masoud Moshref}, \bibinfo{person}{Minlan
  Yu}, \bibinfo{person}{Ramesh Govindan}, {and} \bibinfo{person}{Amin Vahdat}.}
  \bibinfo{year}{2014}\natexlab{}.
\newblock \showarticletitle{Dream: dynamic resource allocation for
  software-defined measurement}. In \bibinfo{booktitle}{\emph{Proceedings of
  the 2014 ACM conference on SIGCOMM}}. \bibinfo{pages}{419--430}.
\newblock


\bibitem[\protect\citeauthoryear{Moshref, Yu, Govindan, and Vahdat}{Moshref
  et~al\mbox{.}}{2015}]%
        {moshref2015scream}
\bibfield{author}{\bibinfo{person}{Masoud Moshref}, \bibinfo{person}{Minlan
  Yu}, \bibinfo{person}{Ramesh Govindan}, {and} \bibinfo{person}{Amin Vahdat}.}
  \bibinfo{year}{2015}\natexlab{}.
\newblock \showarticletitle{Scream: Sketch resource allocation for
  software-defined measurement}. In \bibinfo{booktitle}{\emph{Proceedings of
  the 11th ACM Conference on Emerging Networking Experiments and
  Technologies}}. \bibinfo{pages}{1--13}.
\newblock


\bibitem[\protect\citeauthoryear{Moshref, Yu, Govindan, and Vahdat}{Moshref
  et~al\mbox{.}}{2016}]%
        {moshref2016trumpet}
\bibfield{author}{\bibinfo{person}{Masoud Moshref}, \bibinfo{person}{Minlan
  Yu}, \bibinfo{person}{Ramesh Govindan}, {and} \bibinfo{person}{Amin Vahdat}.}
  \bibinfo{year}{2016}\natexlab{}.
\newblock \showarticletitle{Trumpet: Timely and precise triggers in data
  centers}. In \bibinfo{booktitle}{\emph{Proceedings of the 2016 ACM SIGCOMM
  Conference}}. \bibinfo{pages}{129--143}.
\newblock


\bibitem[\protect\citeauthoryear{Nathan}{Nathan}{1978}]%
        {nathan1978plain}
\bibfield{author}{\bibinfo{person}{Amos Nathan}.}
  \bibinfo{year}{1978}\natexlab{}.
\newblock \showarticletitle{Plain and covariant multivariate fourier
  transforms}.
\newblock \bibinfo{journal}{\emph{Information and Control}}
  \bibinfo{volume}{39}, \bibinfo{number}{1} (\bibinfo{year}{1978}),
  \bibinfo{pages}{73--81}.
\newblock


\bibitem[\protect\citeauthoryear{Oppenheim and Schafer}{Oppenheim and
  Schafer}{1975}]%
        {oppenheim1975digital}
\bibfield{author}{\bibinfo{person}{Alan~V Oppenheim} {and}
  \bibinfo{person}{Ronald~W Schafer}.} \bibinfo{year}{1975}\natexlab{}.
\newblock \showarticletitle{Digital signal processing(Book)}.
\newblock \bibinfo{journal}{\emph{Research supported by the Massachusetts
  Institute of Technology, Bell Telephone Laboratories, and Guggenheim
  Foundation. Englewood Cliffs, N. J., Prentice-Hall, Inc., 1975. 598 p}}
  (\bibinfo{year}{1975}).
\newblock


\bibitem[\protect\citeauthoryear{Owezarski and Larrieu}{Owezarski and
  Larrieu}{2004}]%
        {10.1007/978-3-540-25969-5_9}
\bibfield{author}{\bibinfo{person}{Philippe Owezarski} {and}
  \bibinfo{person}{Nicolas Larrieu}.} \bibinfo{year}{2004}\natexlab{}.
\newblock \showarticletitle{Internet Traffic Characterization -- An Analysis of
  Traffic Oscillations}. In \bibinfo{booktitle}{\emph{High Speed Networks and
  Multimedia Communications}}, \bibfield{editor}{\bibinfo{person}{Zoubir
  Mammeri} {and} \bibinfo{person}{Pascal Lorenz}} (Eds.).
  \bibinfo{publisher}{Springer Berlin Heidelberg}, \bibinfo{address}{Berlin,
  Heidelberg}, \bibinfo{pages}{96--107}.
\newblock
\showISBNx{978-3-540-25969-5}


\bibitem[\protect\citeauthoryear{Penny, Friswell, and Garvey}{Penny
  et~al\mbox{.}}{2003a}]%
        {penny2003letterto}
\bibfield{author}{\bibinfo{person}{JET Penny}, \bibinfo{person}{MI Friswell},
  {and} \bibinfo{person}{SD Garvey}.} \bibinfo{year}{2003}\natexlab{a}.
\newblock \showarticletitle{LETTERTO THE EDITOR DETECTING ALIASED FREQUENCY
  COMPONENTS IN DISCRETE FOURIERTRANSFORMS}.
\newblock \bibinfo{journal}{\emph{Mechanical systems and signal processing}}
  \bibinfo{volume}{17}, \bibinfo{number}{2} (\bibinfo{year}{2003}),
  \bibinfo{pages}{473--481}.
\newblock


\bibitem[\protect\citeauthoryear{Penny, Friswell, and Garvey}{Penny
  et~al\mbox{.}}{2003b}]%
        {PENNY2003473}
\bibfield{author}{\bibinfo{person}{J.~E.~T. Penny}, \bibinfo{person}{M.~I.
  Friswell}, {and} \bibinfo{person}{S.~D. Garvey}.}
  \bibinfo{year}{2003}\natexlab{b}.
\newblock \showarticletitle{DETECTING ALIASED FREQUENCY COMPONENTS IN DISCRETE
  FOURIER TRANSFORMS}.
\newblock \bibinfo{journal}{\emph{Mechanical Systems and Signal Processing}}
  \bibinfo{volume}{17}, \bibinfo{number}{2} (\bibinfo{year}{2003}),
  \bibinfo{pages}{473--481}.
\newblock
\showISSN{0888-3270}
\urldef\tempurl%
\url{https://doi.org/10.1006/mssp.2001.1445}
\showDOI{\tempurl}


\bibitem[\protect\citeauthoryear{Phaal, Panchen, and McKee}{Phaal
  et~al\mbox{.}}{2001}]%
        {sflow}
\bibfield{author}{\bibinfo{person}{Peter Phaal}, \bibinfo{person}{Sonia
  Panchen}, {and} \bibinfo{person}{Neil McKee}.}
  \bibinfo{year}{2001}\natexlab{}.
\newblock \bibinfo{title}{RFC3176: InMon Corporation's sFlow: A Method for
  Monitoring Traffic in Switched and Routed Networks}.
\newblock
\newblock


\bibitem[\protect\citeauthoryear{Romo-C{\'a}rdenas, Avil{\'e}s-Rodr{\'\i}guez,
  S{\'a}nchez-L{\'o}pez, Cos{\'\i}o-Le{\'o}n, Luque, G{\'o}mez-Guti{\'e}rrez,
  Nieto-Hip{\'o}lito, V{\'a}zquez-Brise{\~n}o, and
  Navarro-Cota}{Romo-C{\'a}rdenas et~al\mbox{.}}{2018}]%
        {romo2018nyquist}
\bibfield{author}{\bibinfo{person}{G Romo-C{\'a}rdenas}, \bibinfo{person}{GJ
  Avil{\'e}s-Rodr{\'\i}guez}, \bibinfo{person}{J~de~D S{\'a}nchez-L{\'o}pez},
  \bibinfo{person}{M Cos{\'\i}o-Le{\'o}n}, \bibinfo{person}{PA Luque},
  \bibinfo{person}{CM G{\'o}mez-Guti{\'e}rrez}, \bibinfo{person}{Juan~I
  Nieto-Hip{\'o}lito}, \bibinfo{person}{Mabel V{\'a}zquez-Brise{\~n}o}, {and}
  \bibinfo{person}{Christian~X Navarro-Cota}.} \bibinfo{year}{2018}\natexlab{}.
\newblock \showarticletitle{Nyquist-Shannon theorem application for
  Savitzky-Golay smoothing window size parameter determination in bio-optical
  signals}.
\newblock \bibinfo{journal}{\emph{Results in Physics}}  \bibinfo{volume}{11}
  (\bibinfo{year}{2018}), \bibinfo{pages}{17--22}.
\newblock


\bibitem[\protect\citeauthoryear{Song, Liu, Pang, Hou, and Li}{Song
  et~al\mbox{.}}{2012}]%
        {song2012improved}
\bibfield{author}{\bibinfo{person}{Zhanjie Song}, \bibinfo{person}{Bei Liu},
  \bibinfo{person}{Yanwei Pang}, \bibinfo{person}{Chunping Hou}, {and}
  \bibinfo{person}{Xuelong Li}.} \bibinfo{year}{2012}\natexlab{}.
\newblock \showarticletitle{An improved Nyquist--Shannon irregular sampling
  theorem from local averages}.
\newblock \bibinfo{journal}{\emph{IEEE transactions on information theory}}
  \bibinfo{volume}{58}, \bibinfo{number}{9} (\bibinfo{year}{2012}),
  \bibinfo{pages}{6093--6100}.
\newblock


\bibitem[\protect\citeauthoryear{Stoev, Taqqu, Park, and Marron}{Stoev
  et~al\mbox{.}}{2005}]%
        {STOEV2005423}
\bibfield{author}{\bibinfo{person}{Stilian Stoev}, \bibinfo{person}{Murad~S.
  Taqqu}, \bibinfo{person}{Cheolwoo Park}, {and} \bibinfo{person}{J.S.
  Marron}.} \bibinfo{year}{2005}\natexlab{}.
\newblock \showarticletitle{On the wavelet spectrum diagnostic for Hurst
  parameter estimation in the analysis of Internet traffic}.
\newblock \bibinfo{journal}{\emph{Computer Networks}} \bibinfo{volume}{48},
  \bibinfo{number}{3} (\bibinfo{year}{2005}), \bibinfo{pages}{423--445}.
\newblock
\showISSN{1389-1286}
\urldef\tempurl%
\url{https://doi.org/10.1016/j.comnet.2004.11.017}
\showDOI{\tempurl}
\newblock
\shownote{Long Range Dependent Traffic.}


\bibitem[\protect\citeauthoryear{Tammana, Agarwal, and Lee}{Tammana
  et~al\mbox{.}}{2016}]%
        {tammana2016simplifying}
\bibfield{author}{\bibinfo{person}{Praveen Tammana}, \bibinfo{person}{Rachit
  Agarwal}, {and} \bibinfo{person}{Myungjin Lee}.}
  \bibinfo{year}{2016}\natexlab{}.
\newblock \showarticletitle{Simplifying datacenter network debugging with
  pathdump}. In \bibinfo{booktitle}{\emph{12th $\{$USENIX$\}$ Symposium on
  Operating Systems Design and Implementation ($\{$OSDI$\}$ 16)}}.
  \bibinfo{pages}{233--248}.
\newblock


\bibitem[\protect\citeauthoryear{Tan, Su, Zhang, Lv, Zhang, Miao, Liu, and
  Li}{Tan et~al\mbox{.}}{2021}]%
        {INT}
\bibfield{author}{\bibinfo{person}{Lizhuang Tan}, \bibinfo{person}{Wei Su},
  \bibinfo{person}{Wei Zhang}, \bibinfo{person}{Jianhui Lv},
  \bibinfo{person}{Zhenyi Zhang}, \bibinfo{person}{Jingying Miao},
  \bibinfo{person}{Xiaoxi Liu}, {and} \bibinfo{person}{Na Li}.}
  \bibinfo{year}{2021}\natexlab{}.
\newblock \showarticletitle{In-band network telemetry: a survey}.
\newblock \bibinfo{journal}{\emph{Computer Networks}}  \bibinfo{volume}{186}
  (\bibinfo{year}{2021}), \bibinfo{pages}{107763}.
\newblock


\bibitem[\protect\citeauthoryear{Th{\'e}venaz, Blu, and Unser}{Th{\'e}venaz
  et~al\mbox{.}}{2000}]%
        {thevenaz2000image}
\bibfield{author}{\bibinfo{person}{Philippe Th{\'e}venaz},
  \bibinfo{person}{Thierry Blu}, {and} \bibinfo{person}{Michael Unser}.}
  \bibinfo{year}{2000}\natexlab{}.
\newblock \showarticletitle{Image interpolation and resampling}.
\newblock \bibinfo{journal}{\emph{Handbook of medical imaging, processing and
  analysis}} \bibinfo{volume}{1}, \bibinfo{number}{1} (\bibinfo{year}{2000}),
  \bibinfo{pages}{393--420}.
\newblock


\bibitem[\protect\citeauthoryear{Th{\'{e}}venaz, Blu, and Unser}{Th{\'{e}}venaz
  et~al\mbox{.}}{2000}]%
        {Resampling2000}
\bibfield{author}{\bibinfo{person}{P. Th{\'{e}}venaz}, \bibinfo{person}{T.
  Blu}, {and} \bibinfo{person}{M. Unser}.} \bibinfo{year}{2000}\natexlab{}.
\newblock \showarticletitle{Image Interpolation and Resampling}.
\newblock In \bibinfo{booktitle}{\emph{Handbook of Medical Imaging, Processing
  and Analysis}}, \bibfield{editor}{\bibinfo{person}{I.N. Bankman}} (Ed.).
  \bibinfo{publisher}{Academic Press}, \bibinfo{address}{San Diego CA, USA},
  Chapter~25, \bibinfo{pages}{393--420}.
\newblock


\bibitem[\protect\citeauthoryear{Thompson, Miller, and Wilder}{Thompson
  et~al\mbox{.}}{1997}]%
        {642356}
\bibfield{author}{\bibinfo{person}{K. Thompson}, \bibinfo{person}{G.J. Miller},
  {and} \bibinfo{person}{R. Wilder}.} \bibinfo{year}{1997}\natexlab{}.
\newblock \showarticletitle{Wide-area Internet traffic patterns and
  characteristics}.
\newblock \bibinfo{journal}{\emph{IEEE Network}} \bibinfo{volume}{11},
  \bibinfo{number}{6} (\bibinfo{year}{1997}), \bibinfo{pages}{10--23}.
\newblock
\urldef\tempurl%
\url{https://doi.org/10.1109/65.642356}
\showDOI{\tempurl}


\bibitem[\protect\citeauthoryear{Yaseen, Sonchack, and Liu}{Yaseen
  et~al\mbox{.}}{2018}]%
        {speedlight}
\bibfield{author}{\bibinfo{person}{Nofel Yaseen}, \bibinfo{person}{John
  Sonchack}, {and} \bibinfo{person}{Vincent Liu}.}
  \bibinfo{year}{2018}\natexlab{}.
\newblock \showarticletitle{Synchronized Network Snapshots}. In
  \bibinfo{booktitle}{\emph{Proceedings of the 2018 Conference of the ACM
  Special Interest Group on Data Communication}} (Budapest, Hungary)
  \emph{(\bibinfo{series}{SIGCOMM '18})}. \bibinfo{publisher}{ACM},
  \bibinfo{address}{New York, NY, USA}, \bibinfo{pages}{402--416}.
\newblock
\showISBNx{978-1-4503-5567-4}
\urldef\tempurl%
\url{https://doi.org/10.1145/3230543.3230552}
\showDOI{\tempurl}


\bibitem[\protect\citeauthoryear{Yaseen, Sonchack, and Liu}{Yaseen
  et~al\mbox{.}}{2020}]%
        {tpprof}
\bibfield{author}{\bibinfo{person}{Nofel Yaseen}, \bibinfo{person}{John
  Sonchack}, {and} \bibinfo{person}{Vincent Liu}.}
  \bibinfo{year}{2020}\natexlab{}.
\newblock \showarticletitle{tpprof: A Network Traffic Pattern Profiler}. In
  \bibinfo{booktitle}{\emph{17th {USENIX} Symposium on Networked Systems Design
  and Implementation ({NSDI} 20)}}. \bibinfo{publisher}{{USENIX} Association},
  \bibinfo{address}{Santa Clara, CA}, \bibinfo{pages}{1015--1030}.
\newblock
\showISBNx{978-1-939133-13-7}
\urldef\tempurl%
\url{https://www.usenix.org/conference/nsdi20/presentation/yaseen}
\showURL{%
\tempurl}


\bibitem[\protect\citeauthoryear{Zamaruiev}{Zamaruiev}{2017}]%
        {zamaruiev2017use}
\bibfield{author}{\bibinfo{person}{VV Zamaruiev}.}
  \bibinfo{year}{2017}\natexlab{}.
\newblock \showarticletitle{The use of Kotelnikov-Nyquist-Shannon sampling
  theorem for designing of digital control system for a power converter}. In
  \bibinfo{booktitle}{\emph{2017 IEEE First Ukraine Conference on Electrical
  and Computer Engineering (UKRCON)}}. IEEE, \bibinfo{pages}{522--527}.
\newblock


\bibitem[\protect\citeauthoryear{Zhang, Liu, Zeng, and Krishnamurthy}{Zhang
  et~al\mbox{.}}{2017}]%
        {ubursts}
\bibfield{author}{\bibinfo{person}{Qiao Zhang}, \bibinfo{person}{Vincent Liu},
  \bibinfo{person}{Hongyi Zeng}, {and} \bibinfo{person}{Arvind Krishnamurthy}.}
  \bibinfo{year}{2017}\natexlab{}.
\newblock \showarticletitle{High-Resolution Measurement of Data Center
  Microbursts} \emph{(\bibinfo{series}{IMC '17})}.
  \bibinfo{publisher}{Association for Computing Machinery},
  \bibinfo{address}{New York, NY, USA}, \bibinfo{pages}{78–85}.
\newblock
\showISBNx{9781450351188}


\bibitem[\protect\citeauthoryear{Zhang, Ge, Greenberg, and Roughan}{Zhang
  et~al\mbox{.}}{2005}]%
        {zhang2005network}
\bibfield{author}{\bibinfo{person}{Yin Zhang}, \bibinfo{person}{Zihui Ge},
  \bibinfo{person}{Albert Greenberg}, {and} \bibinfo{person}{Matthew Roughan}.}
  \bibinfo{year}{2005}\natexlab{}.
\newblock \showarticletitle{Network anomography}. In
  \bibinfo{booktitle}{\emph{Proceedings of the 5th ACM SIGCOMM conference on
  Internet Measurement}}. \bibinfo{pages}{30--30}.
\newblock


\bibitem[\protect\citeauthoryear{Zhuo, Ghobadi, Mahajan, F{\"o}rster,
  Krishnamurthy, and Anderson}{Zhuo et~al\mbox{.}}{2017}]%
        {corupt}
\bibfield{author}{\bibinfo{person}{Danyang Zhuo}, \bibinfo{person}{Monia
  Ghobadi}, \bibinfo{person}{Ratul Mahajan}, \bibinfo{person}{Klaus-Tycho
  F{\"o}rster}, \bibinfo{person}{Arvind Krishnamurthy}, {and}
  \bibinfo{person}{Thomas Anderson}.} \bibinfo{year}{2017}\natexlab{}.
\newblock \showarticletitle{Understanding and mitigating packet corruption in
  data center networks}. In \bibinfo{booktitle}{\emph{Proceedings of the
  Conference of the ACM Special Interest Group on Data Communication}}.
  \bibinfo{pages}{362--375}.
\newblock


\end{thebibliography}

\end{document}